\documentclass[fleqn,usenatbib]{mnras}

\usepackage{newtxtext,newtxmath}

\usepackage[T1]{fontenc}
\usepackage{ae,aecompl}

\usepackage{graphicx}	
\usepackage{amsmath}	
\usepackage{amssymb}	
\usepackage{pdflscape}
\title[M33 GMC Catalogue]{A High-Resolution, Dust-Selected Molecular Cloud Catalogue of M33, the Triangulum Galaxy}

\author[T. G. Williams et al.]{
Thomas G. Williams,$^{1}$\thanks{E-mail: thomas.williams@astro.cf.ac.uk}
Walter K. Gear,$^{1}$
and Matthew W. L. Smith$^{1}$
\\
$^{1}$School of Physics \& Astronomy, Cardiff University, Queens Buildings, The Parade, Cardiff, CF24 3AA, UK
}

\date{Accepted XXX. Received YYY; in original form ZZZ}

\pubyear{2018}

\begin{document}
\label{firstpage}
\pagerange{\pageref{firstpage}--\pageref{lastpage}}
\maketitle

\begin{abstract}
We present a catalogue of Giant Molecular Clouds (GMCs) in M33, extracted from cold dust continuum emission. Our GMCs are identified by computing dendrograms. We measure the spatial distribution of these clouds, and characterise their dust properties. Combining these measured properties with CO(\textit{J}=2--1) and 21cm H{\sc i} data, we calculate the gas-to-dust ratio (GDR) of these clouds, and from this compute a total cloud mass. In total, we find 165 GMCs with cloud masses in the range of 10$^4$-10$^7$\,M$_\odot$. We find that radially, $\log_{10}(\mathrm{GDR}) = -0.043(\pm0.038) \,\mathrm{R [kpc]} + 1.88(\pm0.15)$, a much lower GDR than found in the Milky Way, and a correspondingly higher $\alpha_{\rm CO}$ factor. The mass function of these clouds follows a slope proportional to M$^{-2.84}$, steeper than many previous studies of GMCs in local galaxies, implying that M33 is poorer at forming massive clouds than other nearby spirals. Whilst we can rule out interstellar pressure as the major contributing factor, we are unable to disentangle the relative effects of metallicity and H{\sc i} velocity dispersion. We find a reasonably featureless number density profile with galactocentric radius, and weak correlations between galactocentric radius and dust temperature/mass. These clouds are reasonably consistent with Larson's scaling relationships, and many of our sources are co-spatial with earlier CO studies. Massive clouds are identified at large galactocentric radius, unlike in these earlier studies, perhaps indicating a population of CO-dark gas dominated clouds at these larger distances.

\end{abstract}




\section{Introduction}

The study of star-formation and the study of molecular clouds are inexorably linked. As stars are believed to form from the dense molecular gas in these clouds \citep{2010Andre,2010Lada}, our understanding of star-formation is ultimately limited by our ability to resolve ensembles of these star-forming regions. Within our own galaxy, we are faced with the challenges of distance ambiguity -- to overcome this, we can turn to high-resolution mapping of galaxies for studies of large numbers of these molecular clouds. 

One option for locating these molecular clouds is to trace the molecular hydrogen that they contain. However, due to the size and symmetry of the H$_2$ molecule, it is impossible to trace the cold component associated with star formation directly and so a proxy must be employed. Generally, the rotational transitions of CO (the next most common molecule) are favoured, as they are believed to trace the cold molecular gas that resides within these clouds. Resolving these molecular clouds poses a great challenge -- with the average Milky Way (MW) GMC size being $\sim$40\,pc \citep{1979Solomon}, and $\sim$30\,pc in the LMC \citep{2010Hughes}, we are limited to studies in our local Universe (e.g. \citealt{1993Israel,2007Rosolowsky,2010Hughes}). Recently, with the advent of the Atacama Large Millimetre/submillimetre Array (ALMA), these studies can be extended beyond our Local Group of galaxies (e.g. Liu et al. in prep.). 

\begin{table*}
\caption{SCUBA-2 data reduction parameters for both the main data reduction, and the data calibration.}
\label{table:reduction_parameters}
	\begin{tabular}{ccc}
	\hline \hline
	Parameter & Value & Description\\
    \hline
    \multicolumn{3}{c}{Data reduction}\\
    \hline
    maptol & 0.005 & Defines when the map has `converged'.\\
    com.perarray & 0 & Calculate a single common-mode signal for all subarrays.\\
    flt.filt\_edge\_largescale & 120 (450\micron), 320 (850\micron) & Specifies the largest scale structure to be recoverable in the reduction.\\
    ast.zero\_mask & 1 & Provide an external astronomical signal mask, based on the \textit{Herschel} 500\micron \,image.\\
    ast,flt,com.zero\_freeze & 0 & Calculate these masks every iteration.\\
    com.sig\_limit & 5 & Remove high-frequency `blobs' from the map.\\
    flt.filt\_order & 4 & Reduce ringing around bright sources.\\
    flt.ring\_box1 & 0.5 & Reduce ringing around bright sources.\\
    flagslow & 300 & Flag data where sources are obscured by 1/f noise.\\
    \hline
    \multicolumn{3}{c}{Calibration}\\
    \hline
    ast.zero\_mask & 0 & Do not use an external mask.\\
    ast,flt.zero\_circle & 0.033 & Use a circular mask of 120\,arcsec radius.\\
    ast.mapspike & 10 & Ensure very bright pixels are included in the final map.\\
    dcthresh & 10000 & Ensure very bright pixels are included in the final map.\\
    \hline
	\end{tabular}
\end{table*}

Alternatively, an independent method to probe the properties of GMCs uses the cold dust continuum emission of a galaxy. It has long been established that there is a link between the dust content of a galaxy and its molecular gas (e.g. \citealt{1983Hildebrand,2012Magdis,2012Eales}). Thus, the dust continuum allows us an alternative method to CO measurements to probe the properties of GMCs. However, due to the limited resolution of these instruments and the sizes of clouds this method of probing GMCs is only suitable for some of our most nearby galaxies. Using, for example, the \textit{Herschel} Space Observatory \citep{2010Pilbratt}, we can resolve an average-sized molecular cloud up to a distance of around 200kpc at 500\micron \,wavelengths. With the Submillimetre Common-User Bolometer Array 2 (SCUBA-2; \citealt{2013Holland}) on the James Clerk Maxwell Telescope (JCMT), we can resolve these objects up to 600\,kpc away (850\micron), or 1.2\,Mpc (450\micron). However, with ground-based sub-mm observatories we must overcome noise from the sky varying over small scales at the sub-mm wavelengths we probe -- a harsh sky subtraction process must be performed, which has the drawback of also filtering out large-scale structure in these galaxies. Using a Fourier combination technique, we can use space-based observatories operating at similar wavelengths to add this large-scale structure back in to this data, allowing us to retain both the large-scale structure and the much finer structure these ground-based observatories offer. 

M33 provides an excellent laboratory for resolved molecular cloud studies. Located at a distance of 840\,kpc \citep{1991MadoreFreedman}, it is the third massive spiral galaxy of our Local Group, behind our own Milky Way (MW), and Andromeda (M31). Unlike M31, however, M33 is more face-on, with an inclination of 56\textdegree\, \citep{1994ReganVogel}, and so suffers less from projection effects. It is also actively star-forming across its disk \citep{2004Heyer}, and is host to a large number of GMCs. Previous studies of M33 have identified GMCs using line data from $^{12}$CO({\it J}=1--0), such as \cite{1990WilsonScoville}, surveying the inner 2\,kpc of M33 at 7\,arcsec resolution, finding 38 GMCs. All-disk surveys of M33 have suffered from poorer resolution than this, such as \cite{2003Engargiola}, using the {\it J}=1--0 line, and \cite{2012Gratier}, using the {\it J}=2--1 line, finding 148 and 337 GMCs across the disk of M33, respectively. Both of these surveys have resolutions of $\sim$50\,pc, and so many of the GMCs are only marginally resolved.

In this work, we take an alternative approach to map the GMC content of M33. By combining far-infrared and sub-millimetre data, we probe the properties of GMCs via the cold dust continuum emission of M33. The layout of this paper is as follows: we first present an overview of the data used in our study (Sec. \ref{sec:data}), and our method of source extraction (Sec. \ref{sec:gmc_catalogue}). We then move on to measure the properties of these GMCs (Sec. \ref{sec:cloud_properties}) and a comparison to earlier CO surveys (Sec. \ref{sec:comparison_co}). Finally, we summarise our main results (Sec. \ref{sec:conclusions}).

\begin{figure*}
	\includegraphics[width=2\columnwidth]{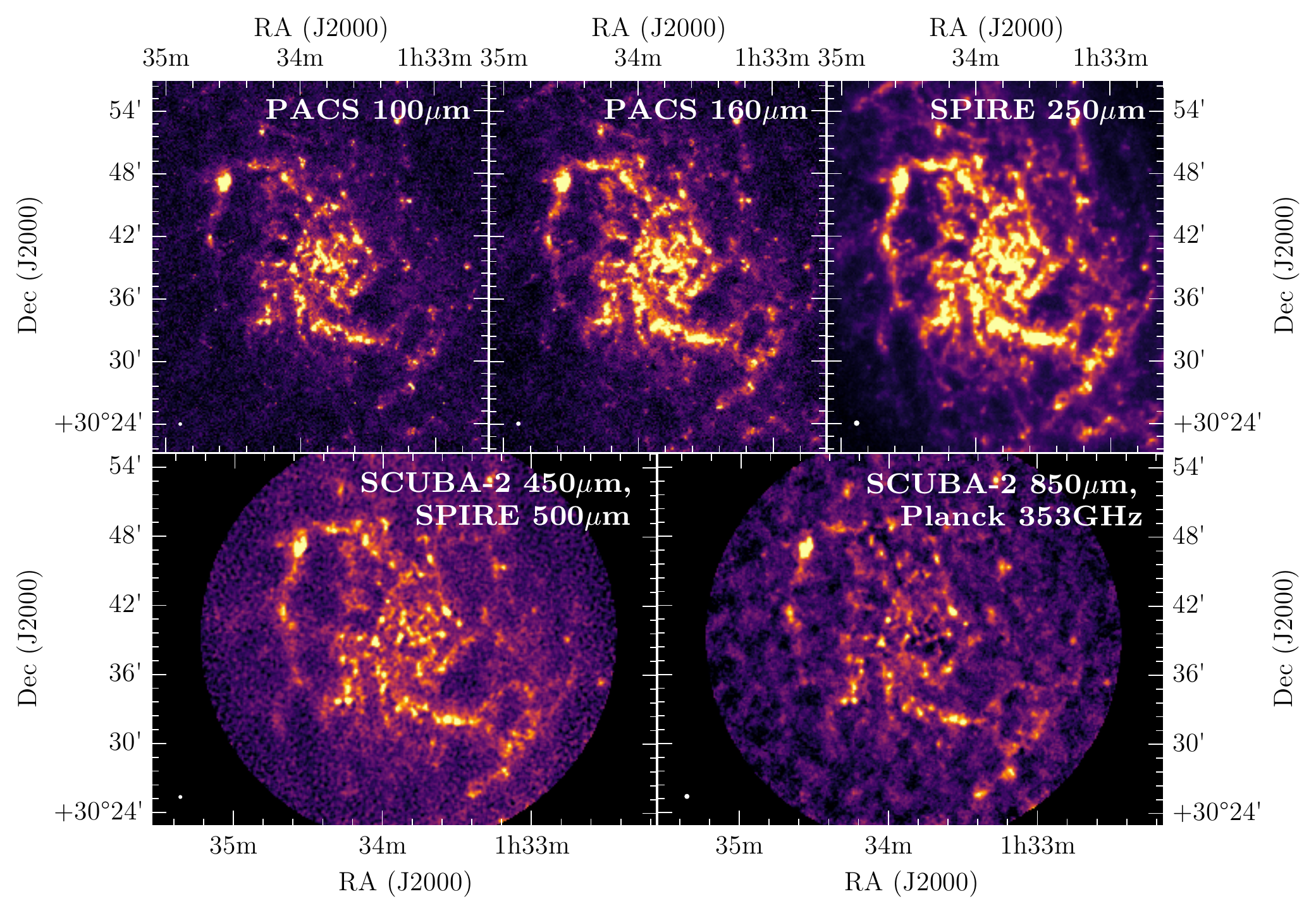}
    \caption{Data used for calculating the dust properties of GMCs in M33. From the top left, PACS 100\micron \,and 160\micron \,data, SPIRE 250\micron \,map, and SCUBA-2 data at 450\micron \,(combined with the SPIRE 500\micron \,map) and 850\micron \,(combined with Planck 353GHz data). To aid with visualisation for the SCUBA-2 maps, we have trimmed to a radius of 15\,arcmin and smoothed slightly with a Gaussian kernel. In each case, the beam is indicated as a white circle in the lower left.}
    \label{fig:dust_data}
\end{figure*}

\section{Data}\label{sec:data}

\subsection{Far Infrared/sub-millimetre}

Our first source of FIR/sub-mm data comes from the \textit{Herschel} Space Observatory. We make use of observations taken as part of the \textit{Herschel} M33 extended survey (HerM33es, \citealt{2010Kramer}), which mapped a 70\,arcmin$^2$ region around M33. Data at 100 and 160\micron \,was taken with the Photoconductor Array Camera and Spectrometer (PACS, \citealt{2010Poglitsch}), with beam sizes of 7.7\,arcsec and 12\,arcsec, respectively. The details of this data reduction are presented in \cite{2011Boquien} and \cite{2015Boquien}. This data has a Root Mean Squared (RMS) noise level of 2.6\,mJy pixel$^{-1}$ (100\micron) and 6.9\,mJy pixel$^{-1}$ (160\micron).

HerM33es simultaneously used the Spectral and Photometric Imaging Receiver (SPIRE, \citealt{2010Griffin}) aboard \textit{Herschel}, which mapped M33 at 250\micron, 350\micron, and 500\micron\, with a resolution of 18\,arcsec, 25\,arcsec, and 36\,arcsec, respectively. This data covers the same region as the PACS maps, to an RMS noise level of 14.1, 9.2, and 8\,mJy\,beam$^{-1}$ at 250, 350 and 500\micron, respectively.

Archival SCUBA-2 observations of M33 at 450 and 850\micron \,were taken between 2012-07-01 and 2012-07-12, consisting of $\sim$7 hours of {\sc pong1800} (which maps a roughly circular, 30 arcmin field) observations of M33, and $\sim$4 hours of smaller, {\sc cv daisy} (constant velocity, small field-of-view) observations. For more details of these SCUBA-2 observing modes, we refer readers to the JCMT observing mode webpage\footnote{\url{https://www.eaobservatory.org/jcmt/instrumentation/continuum/scuba-2/observing-modes/}}. These observations were taken in mostly Band 2/Band 3 weather (225 GHz opacity, $0.04\leq\tau_{225}\leq0.12$). Due to our adopted reduction parameters (see the details on flagslow in Section \ref{s2_data_reduction}), we cannot use these {\sc daisy} maps in our reduction, and so for our purposes, this archival data reaches an RMS noise level of $\sim$6\,mJy\,beam$^{-1}$ at 850\micron, and $\sim$85\,mJy\,beam$^{-1}$ at 450\micron \,(with pixel sizes of 4\,arcsec and 2\,arcsec respectively). As we are particularly interested in the resolution the 450\micron \,data provides, we found that this RMS noise was inadequate and so between 2017-10-17 and 2017-11-21, under Program ID M17BP003 (PI W.K.G.), we obtained a further 12 hours of {\sc pong1800} observations of M33, in good Band 1 weather ($\tau_{225}<0.05$). In the following sections, we describe the data reduction process, which allowed us to create 450\micron \, and 850\micron \,maps of M33 with RMS noise levels of $\sim$35\,mJy\,beam$^{-1}$ and $\sim$4\,mJy\,beam$^{-1}$, respectively. An initial reduction of the data was first presented in \cite{2018Williams}, but in this work we detail this new reduction. The entire dataset used to measure the dust continuum of our GMCs can be seen in Fig. \ref{fig:dust_data}.

The resolution of this SCUBA-2 data is 7.9\,arcsec at 450\micron \,and 13\,arcsec at 850\micron \,\citep{2013Dempsey}, corresponding to 32\,pc and 52\,pc at the distance of M33. However, due to atmospheric variations, extended large-scale structure is filtered out in the reduction process. In order to restore this, we make use of complementary \textit{Herschel} 500\micron \,data for the 450\micron \,data and Planck 353GHz data for the 850\micron \,map. A similar technique has previously been employed with Atacama Pathfinder Experiment Telescope (APEX) Large APEX BOlometer CAmera (LABOCA) data \citep{2016Csengeri} to recover large-scale, extended structure in the Galactic plane, but we have tailored this technique to SCUBA-2 data.

\subsubsection{SCUBA-2 Data Reduction}\label{s2_data_reduction}

The SCUBA-2 data reduction pipeline, {\sc makemap}, is described in detail in \cite{2013Chapin}, and we refer readers to this work for a full description. We used a modified version of this algorithm, called {\sc skyloop}, which performs a single {\sc makemap} iteration each time, including data from all individual observations simultaneously. This helps to constrain the map, and reduce spurious extended emission, which is particularly important for SCUBA-2 observations of local, extended galaxies.

{\sc makemap} is invoked with a file containing the parameters for the map maker. We have attempted to recover some large-scale structure in the SCUBA-2 maps, and so have based our reduction strategy on that of the JCMT Plane Survey (JPS, \citealt{2017Eden}). Our most important, non-default parameters are summarised in Table \ref{table:reduction_parameters} -- for a more detailed description of these parameters, we refer the reader to the SCUBA-2 Data Reduction Cookbook\footnote{\url{http://starlink.eao.hawaii.edu/docs/sc21.htx/sc21.html}}.

\subsubsection{SCUBA-2 Calibration}

{\sc makemap} produces an output file in units of pW, so it is necessary to apply a flux conversion factor (FCF) to the data, to convert it into units of Jy\,beam$^{-1}$. The standard FCFs have been calculated to be 491\,Jy\,beam$^{-1}$\,pW$^{-1}$ at 450\micron, and 537\,Jy\,beam$^{-1}$\,pW$^{-1}$\, at 850\micron \,\citep{2013Dempsey}, but can vary during the night due to effects such as variations in seeing. Particularly for observations near the start of the night, dish cooling can have a major impact on the measured FCF. It is also important to note that the standard FCFs are calculated using a standard configuration file tailored for bright, compact sources, and the configuration parameters can also have an effect. We therefore calibrated our data using FCFs calculated from standard calibrators taken on the same night as the observations. These calibration observations are taken from Mars, Uranus, CRL618, CRL2688, or HL Tau. For observations of M33 between calibrator observations, we take a linear interpolation between the nearest calibrator FCF before and after. In the case that we did not have a calibrator observed either before or after, we took the FCF of the nearest calibrator. We reduced these calibrator observations using the same configuration file as our M33 reduction, with some small modifications (see Table \ref{table:reduction_parameters}). Along with these, we also removed the flagslow parameter, as since these calibration observations are {\sc daisy}s, rather than the larger {\sc pong}s, the telescope was moving slowly enough that all data were flagged.

\begin{figure*}
	\includegraphics[width=2\columnwidth]{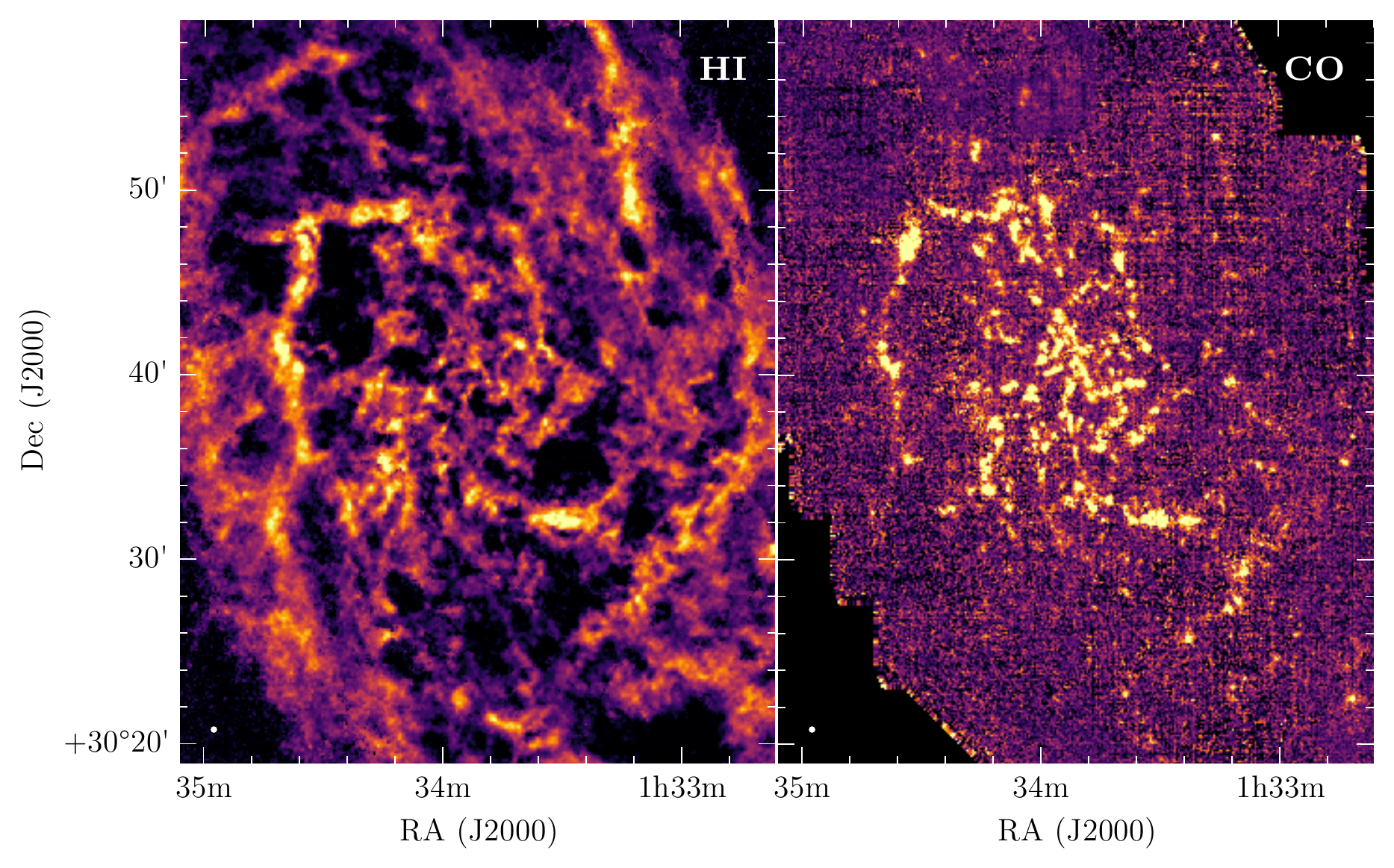}
    \caption{\textit{Left:} 21cm H{\sc i} data and \textit{right:} CO({\it J}=2-1) used in this study. The synthesised beam is indicated as a white circle in the lower left in each case.}
    \label{fig:gas_data}
\end{figure*}

Using this reduction method, we find an average FCF of 522$\pm$51 Jy\,beam$^{-1}$\,pW$^{-1}$ at 450\micron \,(6\% higher than the standard FCF), and 518$\pm$44 Jy\,beam$^{-1}$\,pW$^{-1}$\, at 850\micron \,(4\% lower than the standard FCF). The scatter in FCF is similar to the 10\% at 450\micron \,found by \citep{2013Dempsey}, but higher than the standard 5\% scatter at 850\micron. Having calculated an FCF for each observation, we then multiplied the raw data by the ratio of the calculated to the standard FCF. After then reducing the data using {\sc skyloop}, we multiplied the final map by the standard FCF value. We found that calibrating the data in this way led to an increase in flux of $\sim$3\% in the 450\micron \,map, and a negligible change in the 850\micron \,map compared to simply using the standard FCF on the final map. We also found a decrease in noise of $\sim$3\% in the 450\micron \,map, and $\sim$15\% in the 850\micron \,map.

\subsubsection{Combination with \textit{Herschel} and Planck Data}

As previously mentioned, the SCUBA-2 data reduction process necessarily removes extended structures in the map. However, using a method similar to interferometric `feathering', we can restore this extended structure. Previous work has shown that this technique can work to combine Planck and LABOCA data \citep{2016Csengeri}, but we have tailored this code for SCUBA-2.

First, the units of the two input maps are converted to Jy\,beam$^{-1}$, if necessary. If a SCUBA-2 map is provided in units of pW, the standard FCF is applied. Generally, SPIRE 500\micron \,maps are in units of MJy\,sr$^{-1}$, so we convert to Jy\,beam$^{-1}$ using a beam size of 1665\,arcsec$^2$ (As reported in the SPIRE Handbook\footnote{\url{http://herschel.esac.esa.int/Docs/SPIRE/html/spire_om.html}}). The Planck maps (which are publicly available in HEALPIX format\footnote{\url{https://irsa.ipac.caltech.edu/data/Planck/release_2/all-sky-maps/}}) are provided in units of K$_\mathrm{CMB}$ temperature units, so we convert to Jy\,beam$^{-1}$ using a conversion factor of 287.45\,MJy\,sr$^{-1}$\,K$_\mathrm{CMB}^{-1}$ \citep{2014PlanckIX}, and beam FWHM of 5.19\,arcmin and 4.52\,arcmin \citep{2013PlanckVIII}. We also subtract the contribution of the Cosmic Microwave Background (CMB) from this data following \cite{2015PlanckXXV}, as the CMB varies over scales similar to the extent of M33. We then reproject these maps to the image size and pixel scale of the SCUBA-2 data using Python's {\sc reproject} package.

There are two corrections that must also be applied to the data, to account for the difference in central wavelengths, and colour corrections due to differences in spectral response. In the case of combining SCUBA-2 850\micron \,and Planck 353GHz, the central wavelength correction is negligible. For the \textit{Herschel} data, we perform a central frequency correction, assuming a modified blackbody (MBB), so
\begin{equation}
F(\beta,T) = \left(\frac{500\mu \mathrm{m}}{450\mu \mathrm{m}}\right)^{3+\beta} \times \frac{\exp\left(\frac{\mathrm{hc}}{500\mu \mathrm{m} \times \mathrm{kT}}\right)}{\exp\left(\frac{\mathrm{hc}}{450\mu \mathrm{m} \times \mathrm{kT}}\right)}
\end{equation}
where $\beta$ is the dust emissivity index (if not specified, defaults to 2) and T is the dust temperature (with a default value of 20K).

The colour correction to the Planck data is calculated using
\begin{equation}
C_\mathrm{Planck} = \frac{\int{R(\nu)(\nu/353)^{-1}\mathrm{d}\nu}}{\int{R(\nu)(\nu/353)^{\alpha} \mathrm{d}\nu}}
\end{equation}
where $R(\nu)$ is the Planck 353GHz passband. $\alpha$ is the index of the source spectrum. In the Rayleigh-Jeans spectral regime, $\alpha = 2+\beta$, which gives a default correction factor of 0.854. In the case of the \textit{Herschel} data, we use a factor 1.0049, the colour correction given in Table 5.2 of the SPIRE Handbook for extended sources.

We perform a background subtraction on the Planck and SPIRE 500\micron \,data, using a 3$\sigma$ clipped median. As the SCUBA-2 reduction pipeline models and subtracts the sky, we perform no further sky subtraction on the SCUBA-2 data. Our code applies a Gaussian filter when combining the data, specified by an inputted FWHM. In our case, we set the FWHM to 36\,arcsec for the 450\micron \,data, and 8\,arcmin for the 850\micron \,data. If this value is too small, negative bowling will be present around bright sources, and conversely, if set too high the fine detail desired is lost. We perform Fast Fourier Transforms (FFTs) on the data and the filter to transform them into the uv plane, and create parity between the Jy/beam units by multiplying by the volume ratio of the high- and low-resolution beams. The filter is normalised such that its amplitude at the centre of the uv plane is 1.

The FFT of the low-resolution data is then filtered by multiplying by the FFT of the filter, added to the FFT of the high-resolution data and transformed back into the image plane. There is an inherent uncertainty due to errors in $\beta$ and T, but in practice these are negligible. The total flux density should be determined by the low-resolution map, and we find that the flux density of the low-resolution data alone and the combined data are consistent to well within the calibration uncertainty of the SCUBA-2 data.

We homogenised this dataset to a common resolution (that of the SPIRE 250\micron\, image) and pixel scale. We convolved the data using the method of \cite{2011Aniano}, and regrid to pixel sizes of 6\,arcsec, to ensure that our maps are Nyquist sampled. This regridding is performed using Python's {\sc reproject} routine, which also astrometrically aligns each image.

\subsection{Gas Data}

We also make use of atomic and molecular gas data in this study. H{\sc i} is traced via the 21cm line from archival VLA\footnote{\url{https://science.nrao.edu/facilities/vla/archive/index}} B, C, and D array data (reduced by \citealt{2010Gratier}). The CO data used in this investigation was taken as part of IRAM's M33 Survey Large Program\footnote{\url{http://www.iram.fr/ILPA/LP006/}} \citep{2010Gratier,2014Druard}, which traces the molecular gas out to a radius of 7\,kpc using IRAM's Heterodyne arRAy (HERA, \citealt{2004Schuster}) instrument. This data has an angular resolution of 12\,arcsec and a spectral resolution of 2.6\,km\,s$^{-1}$. These maps can be seen in Fig. \ref{fig:gas_data}.

\section{GMC Catalogue}\label{sec:gmc_catalogue}

\subsection{Identifying GMCs} \label{sec:identifying_gmcs}

\begin{figure*}
	\includegraphics[width=2\columnwidth]{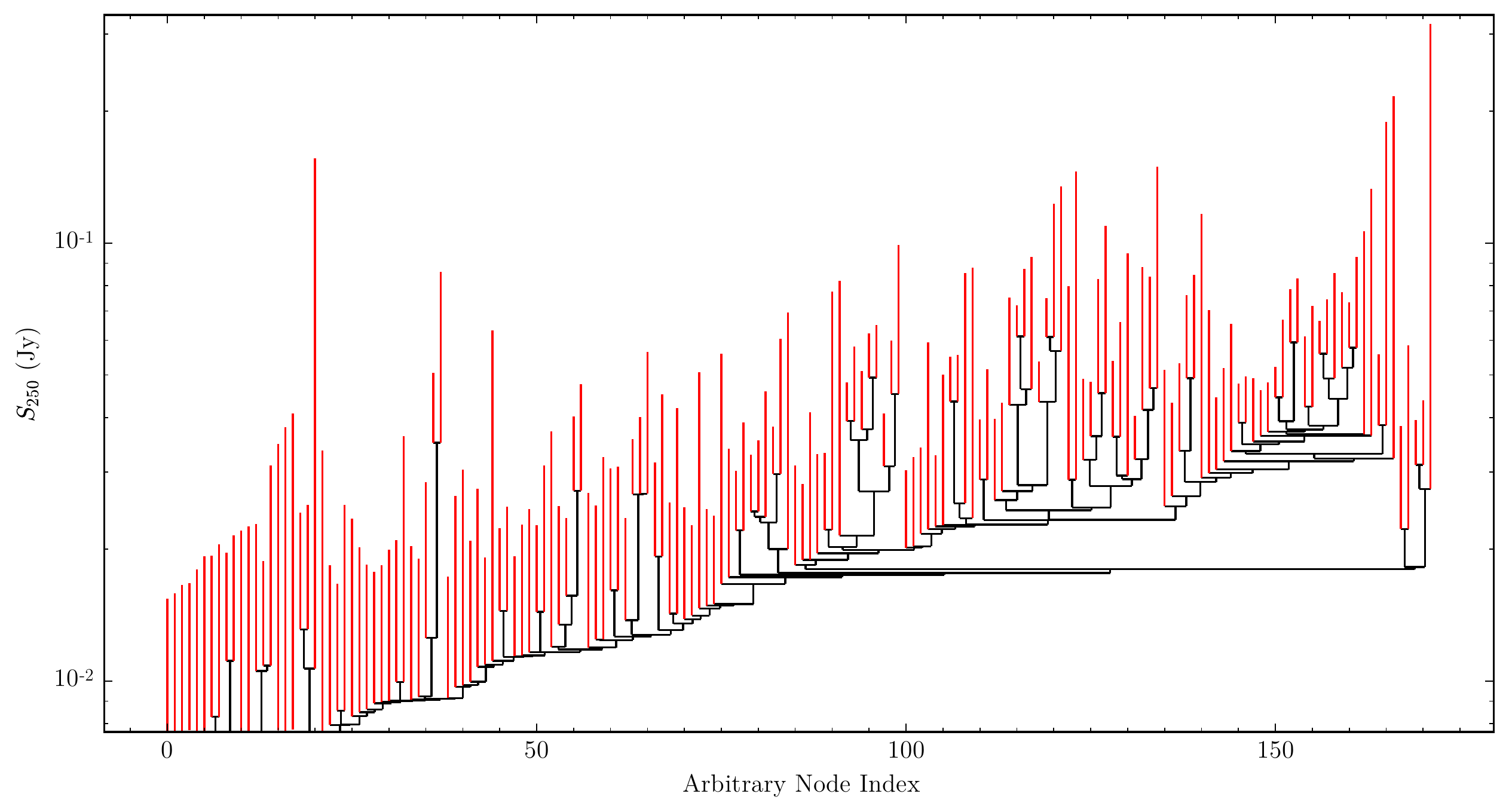}
    \caption{Dendrogram showing SPIRE 250\micron \,flux in M33. The top of each vertical line indicates a leaf node (highlighted in red), which we assume to be our GMCs.}
    \label{fig:dendrogram}
\end{figure*}

\begin{figure*}
	\includegraphics[width=2\columnwidth]{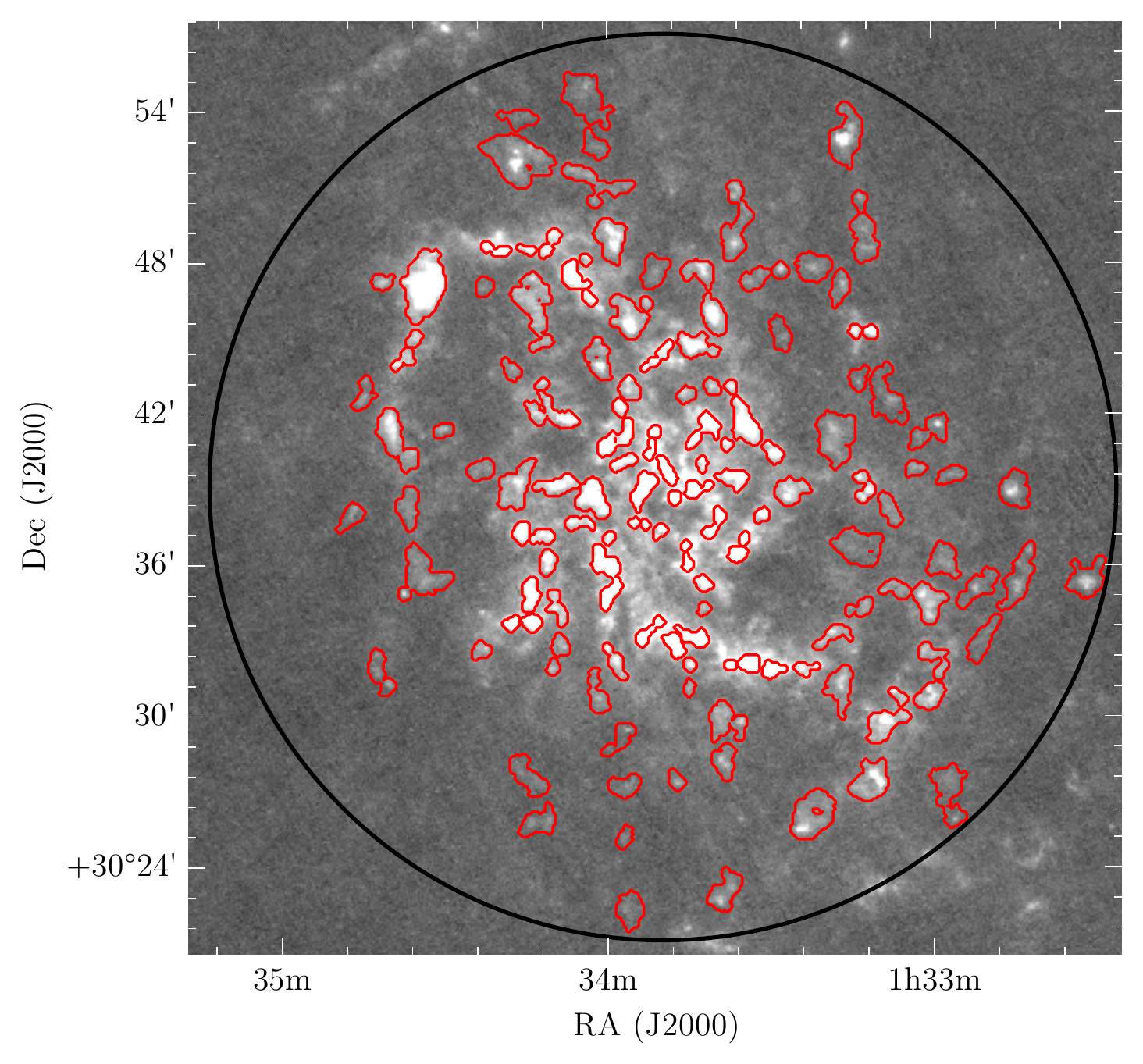}
    \caption{GMC contours (red) overlaid on the PACS160 \micron\,map. The black circle indicates the extent of our search radius, 18\,arcmin.}
    \label{fig:gmc_pos}
\end{figure*}

Disentangling sources from regions of complex emission is a non-trivial task, and several source extraction methods have been developed to achieve this goal (see \citealt{2012Menshchikov} for descriptions of a number of source extraction algorithms). Initial testing using the algorithms {\sc clumpfind} \citep{1994Williams} and {\sc fellwalker} (\citealt{2015Berry}, an algorithm developed to deal with some issues in {\sc clumpfind}) revealed shortcomings in these more traditional methods -- given that much of the emission at these wavelengths is diffuse, the entire galaxy becomes segmented into unreasonably large ``sources''. We also attempted source extraction using SE{\sc xtractor} \citep{1996BertinArnouts}, which can deblend overlapping sources, but this source extraction software only produces an ellipse, and so fails to take into account the irregular nature of many of the structures we are attempting to recover. The structure of a galaxy is hierarchical and interconnected, and so computing a dendrogram of this structure is one way of identifying sources within the galaxy \citep{2008Rosolowsky}; dendrograms also have the additional benefit of extracting nested structure, which is vital in this study. In this work, we use the {\sc astrodendro} dendrogram package\footnote{\url{http://www.dendrograms.org}}. We refer the readers to the documentation on the {\sc astrodendro} website for a more thorough description of the algorithm, but briefly a tree is constructed by arranging the pixels in order of flux. The first structure is centred on the brightest pixel, then the next brightest pixel is checked to see whether it should be considered a new structure or merged into another. The code moves down in flux until neighbouring regions touch, and if the difference between the maxima is significant, these `leaf' structures are merged into a `branch'. The code works down to a minimum value and the structure is complete -- a series of leaves connected to branches, with a `trunk' at the bottom of each structure. These leaves are analagous to traditional sources, and it is these that we consider as our molecular clouds. For a visual comparison of these various algorithms on this data, see Appendix \ref{app:source_extraction_comparison}. We also note that this data does not include kinematic information. This may lead to unrelated, but co-spatial along the line-of-sight, clouds becoming associated to one source when integrating along that line-of-sight. This is highlighted in Sect. \ref{sec:comparison_co}.

As we wish to compute dust properties, we require sufficient data across the dust continuum peak, and into longer wavelengths where the bulk of the mass is contained. We therefore choose five wavebands across this peak, as a balance between spectral coverage and spatial resolution. These are the PACS 100 and 160\micron\, data, the SPIRE 250\micron\, data and the SCUBA-2 450 and 850\micron\, maps. We compute our dendrogram on the SPIRE 250\micron \,data, as we found that after regridding and smoothing to the resolution and pixel scale of our lowest resolution data (the 250\micron), that this map had the highest S/N. We select only regions with flux greater than 3$\sigma$ in each pixel, and regions must have a difference of greater than 3$\sigma$ to be considered significant and separate. This extraction criteria is selected to be as analogous as possible to \cite{2015Kirk}, in order to make our results comparable to this earlier work. We also impose conditions that the region must be bigger than the SPIRE 250\micron \,beam, and that none of these regions can touch the edge of the data. We find 165 leaves (i.e. no resolved substructure) in this dendrogram, which we assume to be our GMCs. The dendrogram for M33 can be seen in Fig. \ref{fig:dendrogram}, and the positions of these clouds in Fig. \ref{fig:gmc_pos}. The majority of our analysis was performed on these clouds, although we also highlight the effect of performing this extraction on the SCUBA-2 450\micron\, map (our highest resolution data) in the size distribution (Sect. \ref{sec:size_distribution}).

\begin{figure*}
	\includegraphics[width=2\columnwidth]{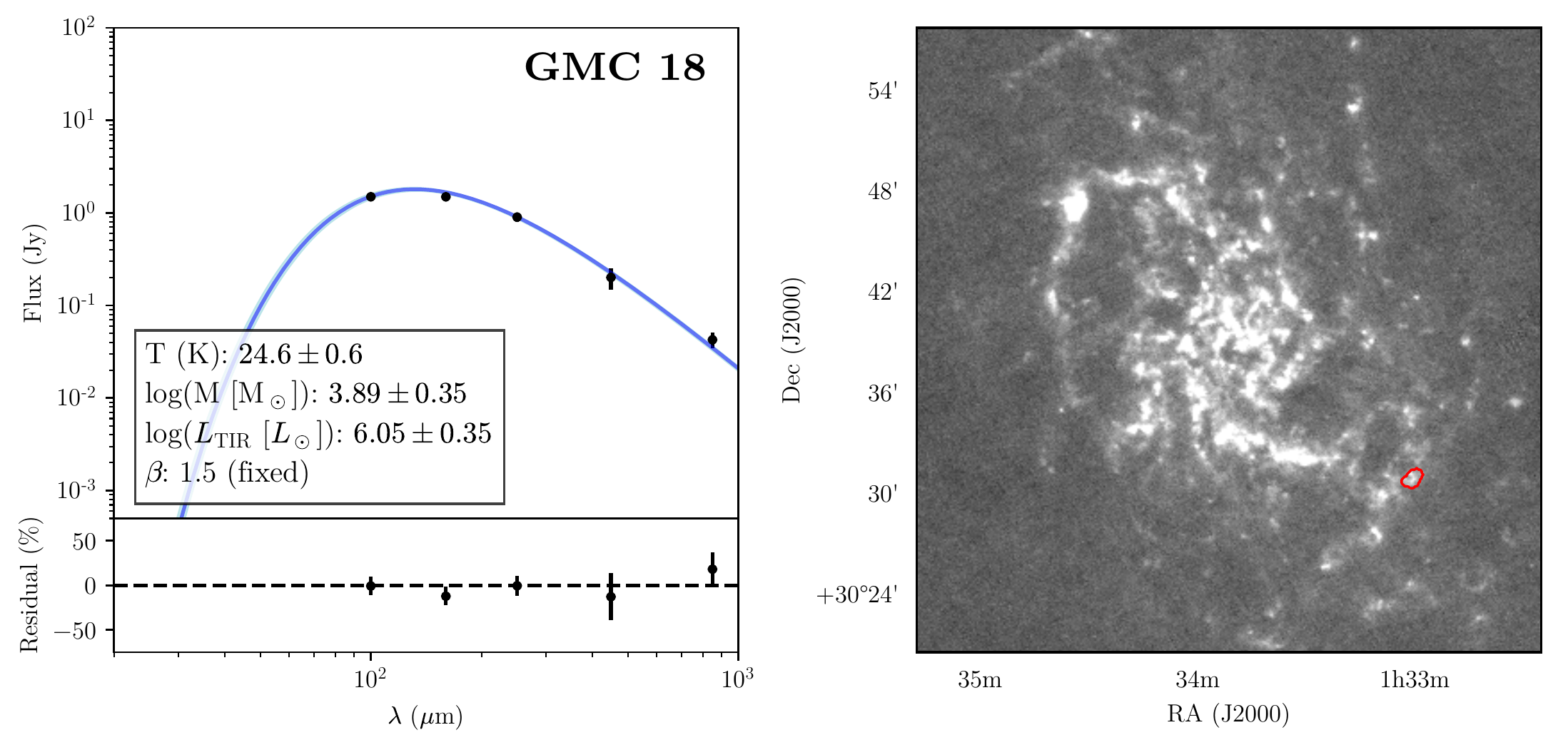}
    \caption{An example SED, from GMC ID 18. The left panel shows the fit, along with various parameters. The 1$\sigma$ error is shown as the shaded region (not including error in $\kappa_{500}$). The right shows the contour of this cloud overlaid on the PACS 160\micron \,map.}
    \label{fig:example_sed}
\end{figure*}

\subsection{Flux Extraction and SED Fitting} \label{sec:flux_extract_sed_fit}

Various parameters of these leaf nodes can be seen in Table \ref{table:leaf_node_parameters}. For each node, we list the mean position of the structure, and the deprojected distance from the centre of M33 (01h33m50.9s, +30$\degr$39$^\prime$37$^{\prime\prime}$; \citealt{2008Plucinsky}). We also calculate a FWHM of the cloud, based on its intensity-weighted second moment.

We have also computed fluxes in the PACS and SCUBA-2 bands for each of these clouds. {\sc astrodendro} outputs a mask for each node, and we measured the flux beneath each mask in each waveband for all of the nodes. We estimated a local background from the median of the isocontour surrounding the mask, which we subtracted from the pixels before summation. An estimate of the local RMS error is given by the standard deviation of the pixels in this isocontour. The fluxes are also listed in Table \ref{table:leaf_node_parameters} -- for fluxes less than 3$\sigma$, we list the flux as an upper limit. The error listed in this table reflects only the RMS error; we have not included any calibration error in this value.

For the clouds with fluxes > $3\sigma$ in 3 or more bands (in our case, every cloud), we fit a single modified blackbody (MBB) of the form
\begin{equation}\label{eq:mbb}
S_\nu = \frac{\kappa_\nu {\rm M}_\text{dust}B(\nu,T_\text{dust})}{D^2},
\end{equation} 
where $S_\nu$ is the flux at frequency $\nu$, $\kappa_\nu$ is dust absorption coefficient at frequency $\nu$, i.e.
\begin{equation}\label{eq:kappa}
\kappa_\nu = \kappa_{\nu_0} \left( \frac{\nu}{\nu_0} \right)^\beta,
\end{equation} 
M$_\text{dust}$ is the dust mass, $B(\nu,T_\text{dust})$ is the Planck function at frequency $\nu$ and dust temperature $T_\text{dust}$, and D is the distance to the source. We normalise $\kappa_\nu$ using the value calculated by \cite{2016Clark}, $\kappa_{500} = 0.051^{+0.070}_{-0.026} \,\mathrm{m}^2 \,\mathrm{kg}^{-1}$. We note that this only holds true for the optically thin case, but as the theoretically expected value for when the optical depth becomes unity is 100\micron \,\citep{2006Draine}, and experimentally only affects points shorter than 50\micron\,\citep{2012Casey}, this is a reasonable assumption for our fits. To minimise the number of free parameters in this fit, we assumed a fixed $\beta$ of 1.5, which \cite{2014Tabatabaei} find to be a good fit to M33. We include correlated uncertainties in the PACS and SPIRE bands (as the SCUBA-2 450\micron\, data includes the SPIRE 500\micron\, map). This is implemented by employing the full covariance matrix. We performed our fitting within an MCMC framework using {\sc emcee}\footnote{\url{http://dfm.io/emcee/current/}}, and we quote the errors as the 84$^{\rm th}$ percentile minus the 50$^{\rm th}$ percentile, as we find that our errors are symmetric. Our initial guess for dust mass and temperature were set from a simple least-squares fit. We allowed the dust temperature to freely vary from 0 to 200K, and the dust mass from 0 to 10$^{13}$ M$_\odot$. An example SED fit is shown in Fig. \ref{fig:example_sed}. We also calculated the total infrared (TIR) luminosity of this cloud by integrating the MBB from 3-1100\micron. We find that these clouds contribute around 50\% of the total dust luminosity of M33, despite only occupying around 20\% of the area. This indicates that these clouds are, in general, compact and bright in their dust emission. All of our derived SED parameters are given in Table \ref{table:sed_parameters}. The dominant error in the dust mass and luminosity is error in $\kappa_{500}$ ($\sim$0.32\,dex). As this is a systematic error we do not include this in Table \ref{table:sed_parameters}. We do, however, include this uncertainty in our analysis.

We also include a measurement of the CO({\it J}=2--1) luminosity (in K km\,s$^{-1}$) in Table \ref{table:sed_parameters}. Finally, we calculated H{\sc i} surface densities (in M$_\odot$\,pc$^{-2}$) of each of our sources. A surface density is calculated, assuming \citep{1996Rohlfs}
\begin{equation}
\Sigma_\text{H{\sc i}} = 1.8\times10^{18} \,\text{cm}^{-2}/(\text{K km s}^{-1}).
\end{equation}
With this gas data, we performed the same procedure as for the FIR/sub-mm flux extraction -- convolution and regridding to the same pixel scale, as well as local background subtraction. Similarly to the FIR/sub-mm fluxes, we list upper limits for intensities less than 3$\sigma$.

\section{Cloud Properties}\label{sec:cloud_properties}

\subsection{Size Distribution}\label{sec:size_distribution}

\begin{figure}
	\includegraphics[width=\columnwidth]{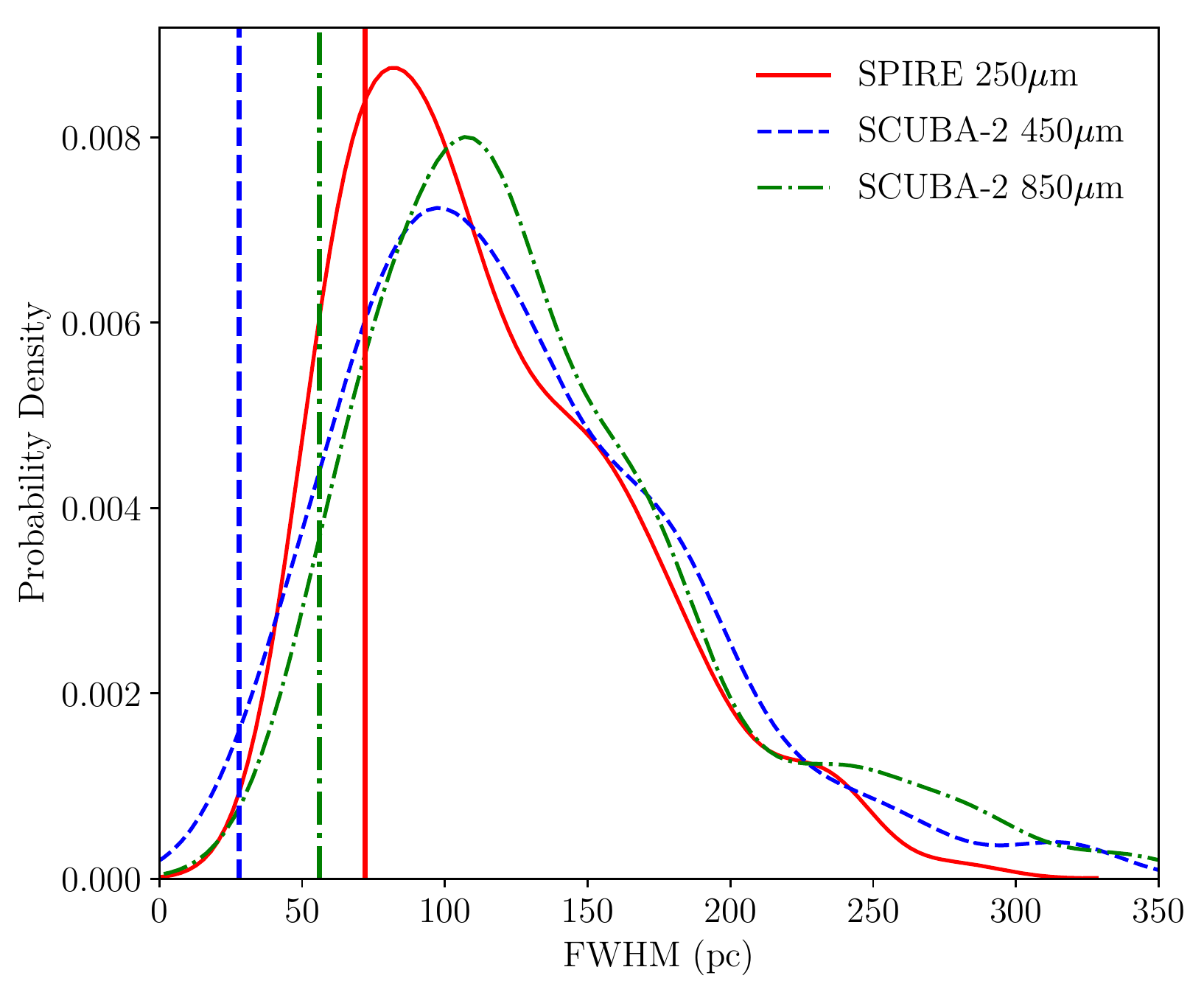}
    \caption{Kernel Density Estimator plot for size distribution of GMCs in M33. The red solid indicates the distribution of the SPIRE 250\micron\, sources, the blue dashed line the SCUBA-2 450\micron\, sources, the green  dot-dashed line the SCUBA-2 850\micron. The equivalently coloured vertical line shows the beam size for the particular instrument (which we enforce our clouds to be larger than).}
    \label{fig:size}
\end{figure}

\begin{figure}
	\includegraphics[width=\columnwidth]{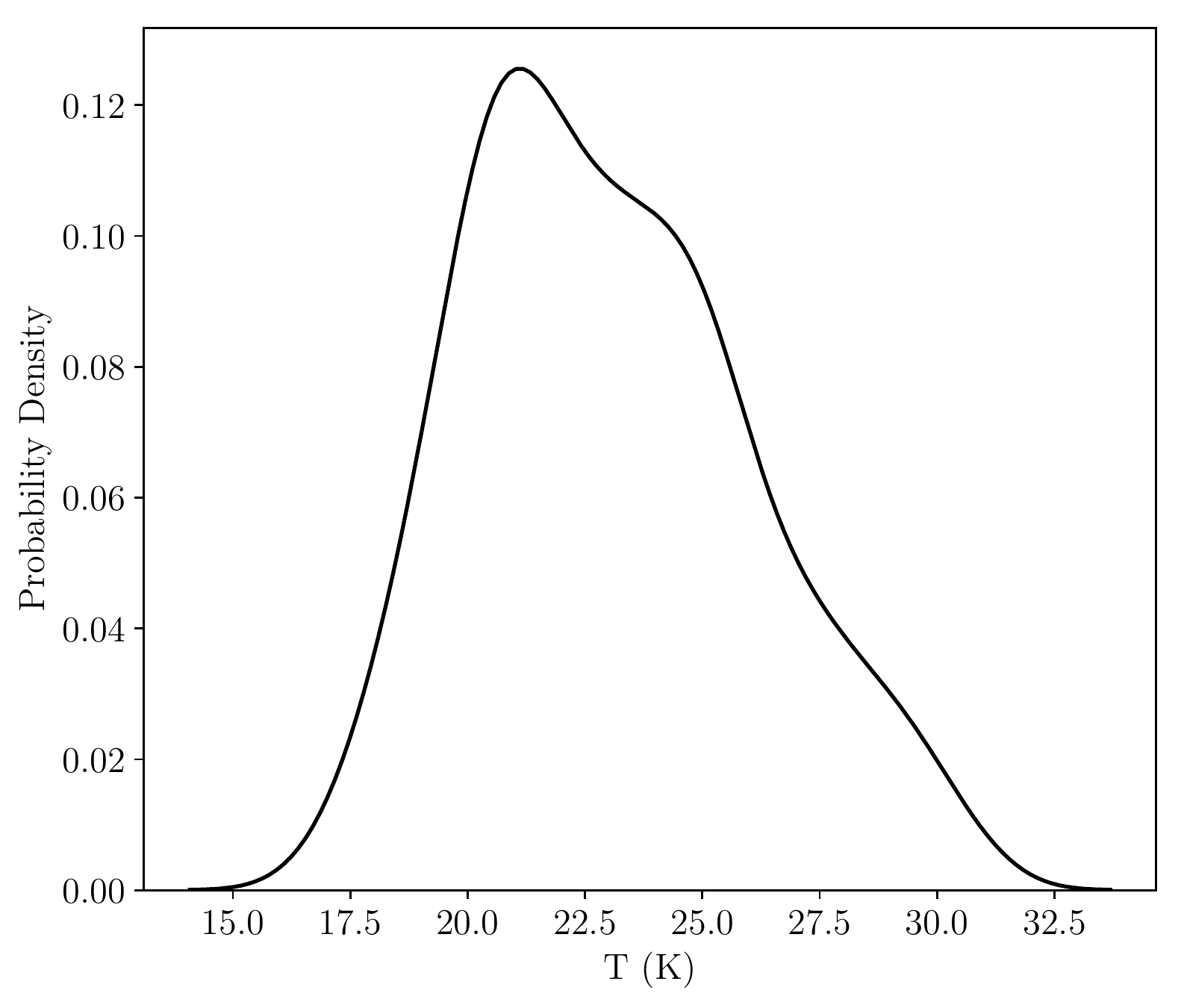}
    \caption{Kernel Density Estimator plot for temperature distribution of GMCs in M33.}
    \label{fig:t_distribution}
\end{figure}

For each source, we take the ellipse enclosing the cloud as computed by {\sc astrodendro} from the half-width at half maximum (HWHM) of the second moments. We calculate a FWHM for each cloud from the average of these HWHM. The size distribution of the clouds can be seen as a Kernel Density Estimator (KDE) plot in Fig. \ref{fig:size}. The median size of these clouds is 105\,pc, close to the FWHM of the SPIRE 250\micron \,beam, and so may initially be assumed to be complexes of smaller clouds. However, when performing the same extraction on our higher-resolution SCUBA-2 450\micron\, and 850\micron\, data, which has a minimum FWHM of 28 and 56\,pc as defined in our extraction criteria, very similar trends are seen (Fig. \ref{fig:size}). This would indicate that these objects are either (a) genuinely more extended than seen in the MW or (b) complexes of many very small clouds, rather than several larger clouds. \cite{2010Roman-Duval} find cloud sizes of 0.2 to 35pc, with a mean size of $\sim$8pc. More recently, \cite{2017Miville-Deschenes} find MW cloud sizes up to $\sim500$pc, with a mean size of $\sim$30pc. Given these results from the MW, this would indicate these sources are likely complexes of smaller clouds. Additionally, comparisons to CO surveys (see Sect. \ref{sec:comparison_co}) show that scenario (b) is more likely the case.

\subsection{Dust Temperatures}

One of the fitted parameters in the MBB is the dust temperature, and the distribution of this is shown in Fig \ref{fig:t_distribution}. We find that the clouds have a median temperature of $23\pm4$K, somewhat warmer than found for clouds in M31, which have a dust temperature of $18\pm2$K \citep{2015Kirk}. We attribute this to the fact that M33 is much more actively star-forming than M31 \citep{2004Heyer}, and thus this dust is more strongly irradiated by these young stars. The distributions of size (Fig. \ref{fig:size}) and dust temperature (Fig. \ref{fig:t_distribution}) look somewhat similar. However, a calculation of the Kendall rank correlation coefficient \citep{1938Kendall}, where $\tau=+1$ indicates a perfect correlation, and $\tau=-1$ a perfect anti-correlation gives a weak anti-correlation of -0.17. A two-sample Kolmogorov-Smirnov test gives a p-value  $\ll$1\%, indicating that these two distributions are significantly different.

\subsection{Cloud Masses}

\subsubsection{Calculating Masses}

\begin{figure}
	\includegraphics[width=\columnwidth]{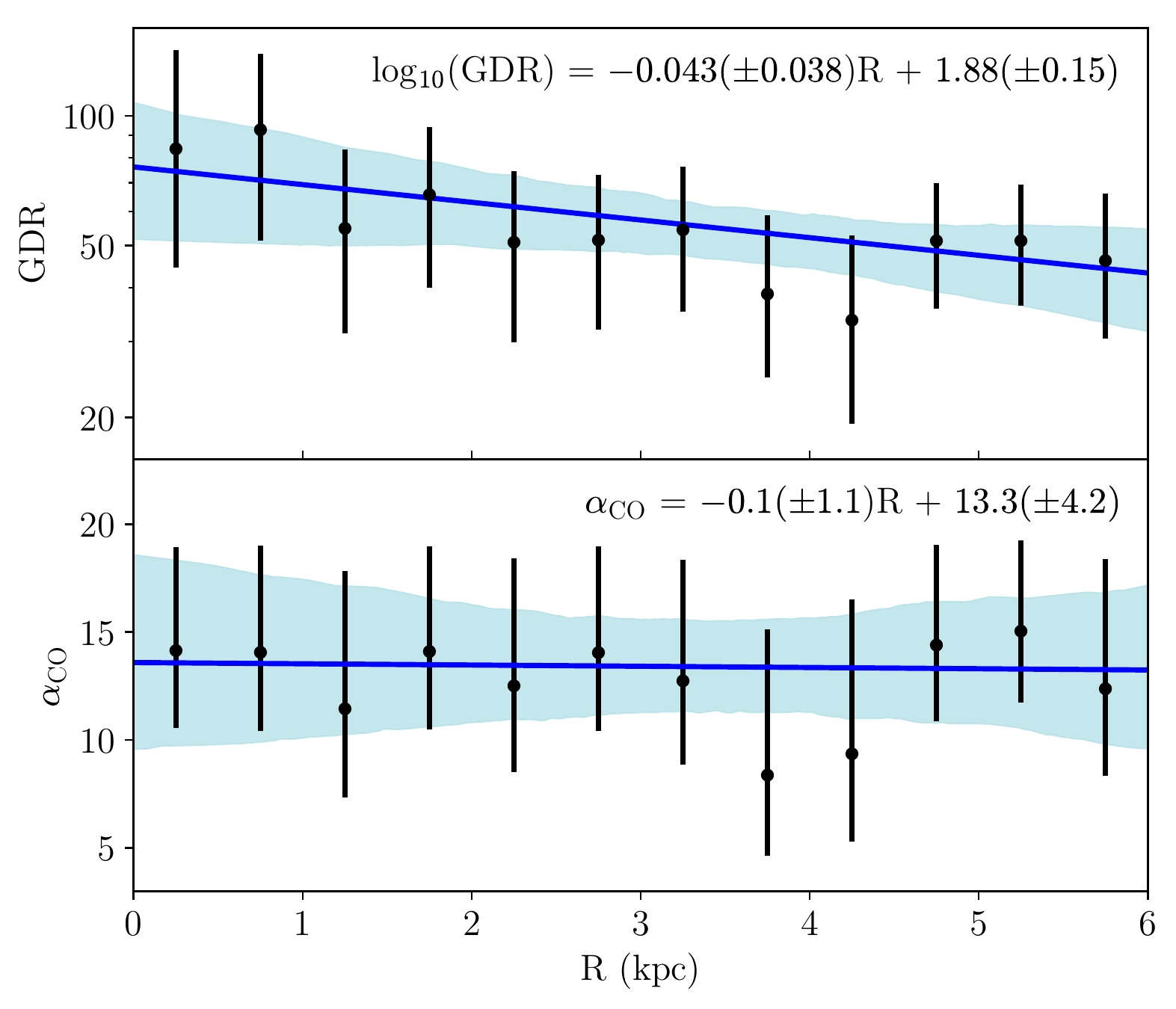}
    \caption{{\it Top:} radial variation in the GDR of M33. The blue line shows the median fit to the data, the blue shaded region the 1$\sigma$ errors on this fit. {\it Bottom:} radial variation in $\alpha_{\rm CO}$. The line and shaded region have the same meanings as the top panel.}
    \label{fig:radial_gdr}
\end{figure}

\begin{figure*}
	\includegraphics[width=2\columnwidth]{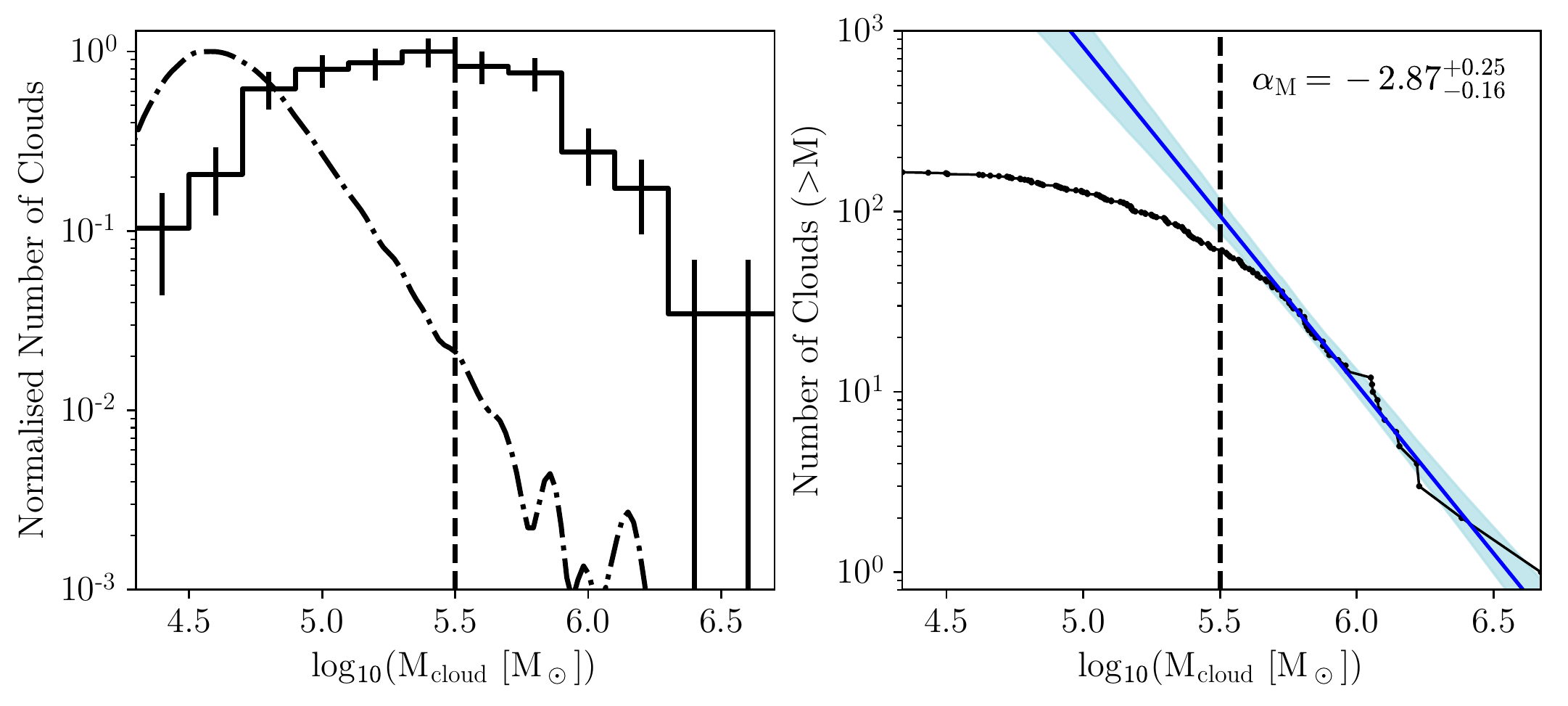}
    \caption{\textit{Left:} Cloud mass distribution for clouds in M33. The dot-dash line shows the point source sensitivity. The vertical dashed line shows the 95\% completeness limit. Both of these distributions are normalised such that the peak is unity. \textit{Right:} Cumulative cloud mass distribution for clouds in M33. The vertical dashed line again represents the 95\% completeness limit, above which we fit the power-law relationship. The blue line shows the power-law fit to the data points above $10^{5.5}$M$_\odot$. 1$\sigma$ errorbars on this power-law fit are shown as the blue shaded region.}
    \label{fig:gmc_mass_function}
\end{figure*}

We simultaneously calculate a gas-to-dust ratio (GDR) and CO conversion factor ($\alpha_\mathrm{CO}$) in a fashion similar to that of \cite{2013Sandstrom}. A dust mass surface density can be converted to a total gas mass surface density via
\begin{equation}
\mathrm{GDR} \times \Sigma_\mathrm{Dust} = \Sigma_\mathrm{HI} + \alpha_\mathrm{CO} \times I_\mathrm{CO}.
\end{equation} 
\cite{2013Sandstrom} find the best fit of these two unknown parameters simultaneously by minimising the scatter in the log of the dust to gas ratio (DGR), and we perform this fitting using an MCMC analysis, accounting for errors in the dust mass surface density, H{\sc i} surface density and CO intensity.

We performed this fitting by grouping our clouds into bins of increasing galactocentric radius. We split our clouds into radial bins of 0.5\,kpc and simultaneously fit $\alpha_\mathrm{CO}$ and GDR. The results of this can be seen in Fig. \ref{fig:radial_gdr}, and the slope for the GDR is given by
\begin{equation}\label{eq:gdr_eqn}
\log_{10}(\mathrm{GDR}) = -0.043(\pm0.038) \,\mathrm{R [kpc]} + 1.88(\pm0.15).
\end{equation}
Our maximum GDR is somewhat lower than seen in nearby galaxies \citep{2013Sandstrom}, with a value of around 90. However, variation in $\kappa_\nu$ can easily lead to huge variations in dust mass. Given that our adopted $\kappa_\nu$ is on the low end of literature values \citep{2016Clark}, this is not unexpected. However, we note that whilst our adopted $\kappa_\nu$ is lower than many other literature estimates, it is still compatible with the $\kappa_\nu$ of \cite{2007DraineLi}, which \cite{2013Sandstrom} use in their work. Thus, this low value for the GDR cannot simply be attributed to our choice of $\kappa_\nu$. Using this calculated GDR, we transformed the dust mass into a total gas mass using Equation \ref{eq:gdr_eqn} and then calculated a total cloud mass (the sum of the dust and gas mass). The mass distribution can be seen in the left panel of Fig. \ref{fig:gmc_mass_function}.

We also calculated $\alpha_{\rm CO}$ simultaneously radially, and we show this in the bottom panel of Fig. \ref{fig:radial_gdr}. This is a CO conversion factor for the {\it J}=1--0 line, assuming CO(2--1)/CO(1--0) = 0.7 \citep{2013Sandstrom}. There is little radial variation, unlike the GDR, but the value is much higher than seen in other, nearby galaxies \citep{2013Sandstrom}. Even given variation in $\kappa_\nu$ that could decrease these values by a factor $\sim$2, this would indicate a CO conversion factor, $\alpha_{\rm CO}$, that is higher than seen in other, nearby galaxies. This is likely due to the subsolar metallicity of M33, with CO molecules becoming more easily dissociated by UV radiation at low metallicity \citep{2016GloverClark}.

We estimated the point source mass sensitivity by taking a limiting flux of 68.9mJy (a 3$\sigma$ point source as defined by our dendrogram extraction criteria), and sampled the GDR and dust temperature from distributions given by the distributions of our clouds (T = $23\pm4$K, log(GDR) = $1.74\pm0.09$). Bootstrapping this 10,000 times, we find a point-source sensitivity of $4.63^{+0.28}_{-0.22} \,\log_{10}(\mathrm{M}_\odot)$. This is shown in Fig. \ref{fig:gmc_mass_function}, and cannot fully account for the deviation from a power-law at the low-end of the mass distribution.

We next estimated the completeness by injecting point sources of given cloud mass into a fake map with the same noise properties as the SPIRE 250\micron\, data, and a background similar to that of M33. We sample the dust temperature and GDR as with the point source sensitivity, and inject 100 sources of each mass into this map. We performed the same extraction criteria as we did with the real data and calculated the completeness for each mass. We find that we are 95\% complete above a mass of 10$^{5.5}$ M$_\odot$. This means that the observed downturn is simply due to incompleteness, and is {\it not} a genuine turnover. However, we must stress that this is only an approximation of the true completeness limit. We have here assumed only point sources present in a constant background, but given that these sources are extended, and embedded in a complex background, the true completeness limit will be a function of mass, radius, cloud shape and position within the map. Accounting for this complex completeness is beyond the scope of this work.

\subsubsection{Power-Law Fitting}

We fit a power-law of the form $N$ (M) $\propto$ M$^{\alpha_{\rm M}}$ to the high-end of the mass distribution. However, in a standard distribution the fit can become biased by small number statistics at high-mass \citep{2009Maschberger}, and so it is more reliable to fit to the cumulative mass distribution (shown in the right panel of Fig. \ref{fig:gmc_mass_function}). In this case, the power-law takes the form $N (>{\rm M}) \propto {\rm M}^{\alpha_{\rm M}-1}$. To avoid incompleteness, we fit only to values with a cloud mass greater than $10^{5.5}$M$_\odot$. We find a value of $\alpha_{\rm M}$ of $-2.83^{+0.24}_{-0.15}$, steeper than the value of $\alpha_{\rm M} = -2.0\pm0.1$ found previously in M33 by \cite{2012Gratier} using CO({\it J}=2--1), and $-2.6\pm0.3$ from the CO({\it J}=1--0) work of \cite{2003Engargiola}. Work by \cite{2010Bigiel} has hinted at a steeper slope in the outskirts of M33, and our calculated slope appears to confirm this. We also find that this value is steeper than molecular clouds in the MW, which has an exponent of around -1.5 (e.g. \citealt{1985Sanders,1979Solomon}). The slope is also steeper than that found in M31 ($-2.34\pm0.21$, \citealt{2015Kirk}, $-2.55\pm0.2$, \citealt{2007Blitz}). The steepness of this slope appears to indicate that M33 is more dominated by smaller clouds than in, e.g., the MW. Given that \cite{2012Gratier} use the CO luminosity as a proxy for molecular hydrogen, whilst the dust content should be an independent tracer of total gas content, we can rule out this steep slope being due to a lower CO intensity per H$_2$. It would appear that M33 is intrinsically poorer at cloud assembly than other local spirals.

The efficiency of cloud assembly has been linked to a variety of processes. The amplitude of the spiral density wave can have an effect on the GMC population (e.g. \citealt{1972Shu}). However, given that recent modelling work has shown that the spiral arms of M33 are most likely driven by gravitational instabilities \citep{2018Dobbs}, we believe this is unlikely to be the case. The interstellar pressure of gas \citep{1994ElmegreenParravano,2006BlitzRosolowsky} may also be a factor. However, we see higher interstellar pressure in M33 than in the MW \citep{2008KasparovaZasov}, so given this hypothesis we would expect more massive clouds. We can therefore rule out the interstellar pressure as the main driver of this inefficient cloud formation. Alternatively, metallicity can play a role in the conversion of H{\sc i} to H$_2$ \citep{2008Krumholz}. Given the subsolar metallicity of M33, we would expect this conversion to be less efficient, and therefore cloud formation similarly inefficient. Finally, it is believed that H$_2$ can form from merging H{\sc i} clouds (e.g. \citealt{2005Heitsch}), so we may expect from larger H{\sc i} velocity dispersions, more massive clouds may form. The average H{\sc i} velocity dispersion in M33 is of the order 13\,km\,s$^{-1}$, with little radial variation \citep{2018Corbelli}, whilst the outer MW shows much more turbulent H{\sc i} gas, with velocity dispersions of 74\,km\,s$^{-1}$ \citep{2008Kalberla}. Our results are unable to distinguish which of these two mechanisms are the main driving force behind this inefficient cloud formation, but it is clear that the cloud mass distribution is significantly different in M33 than the other massive spirals in our Local Group. The exact cause of this is currently unclear, but high-resolution surveys of many galaxies with a wide range of properties will be able to explain the diversity in cloud populations seen even between the galaxies of our Local Group.

\subsection{Radial Variation in Cloud Properties}

\begin{figure}
	\includegraphics[width=\columnwidth]{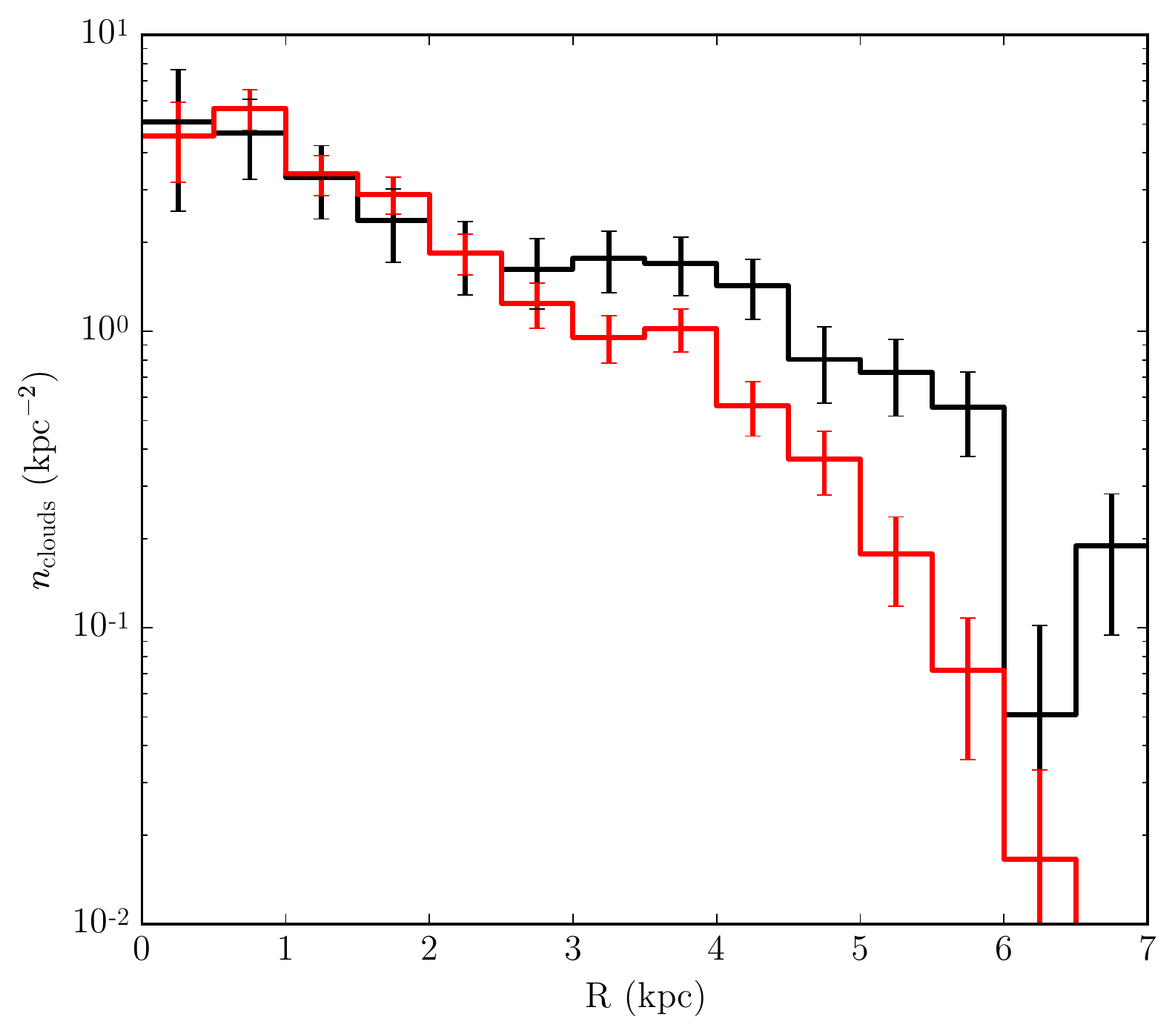}
    \caption{Number density of GMCs with galactocentric radius. The black line indicates the clouds in our study, the red from the work of \citet{2012Gratier}. The curves are normalised at the 2-2.5\,kpc bin.}
    \label{fig:number_density}
\end{figure}

We also investigated any radial variation in the cloud properties. Fig. \ref{fig:number_density} shows the number density (the number of clouds per annular area) of these clouds with galactocentric radius. We see that up to a radius of 2.5\,kpc, our cloud distribution and the GMC distribution of \cite{2012Gratier} agree very well -- however, after $\sim$3\,kpc, the distribution of GMCs from \cite{2012Gratier} is systematically lower. We believe that this is due to the fact that our data is wider than the CO map on which they perform their extraction. As they perform this analysis on an incomplete map of M33 (only the area covered by the \textit{Herschel} PACS and Heterodyne Instrument for the Far Infrared (HIFI) spectrometers), they do not map the entire disk of M33. We would suggest our distribution is therefore less biased, and gives a more representative view of the GMC number density. Along with a peak in the distribution at the centre of the galaxy, there is a step in this distribution from around 2\,kpc to 4\,kpc radius, which corresponds to the positions of the spiral arms in M33. However, the spiral arms are less pronounced in this distribution than M31 \citep{2015Kirk}, where the positions of the spiral arms have clear peaks, and the SFR in the rest of M31 is very low.

\begin{figure}
	\includegraphics[width=\columnwidth]{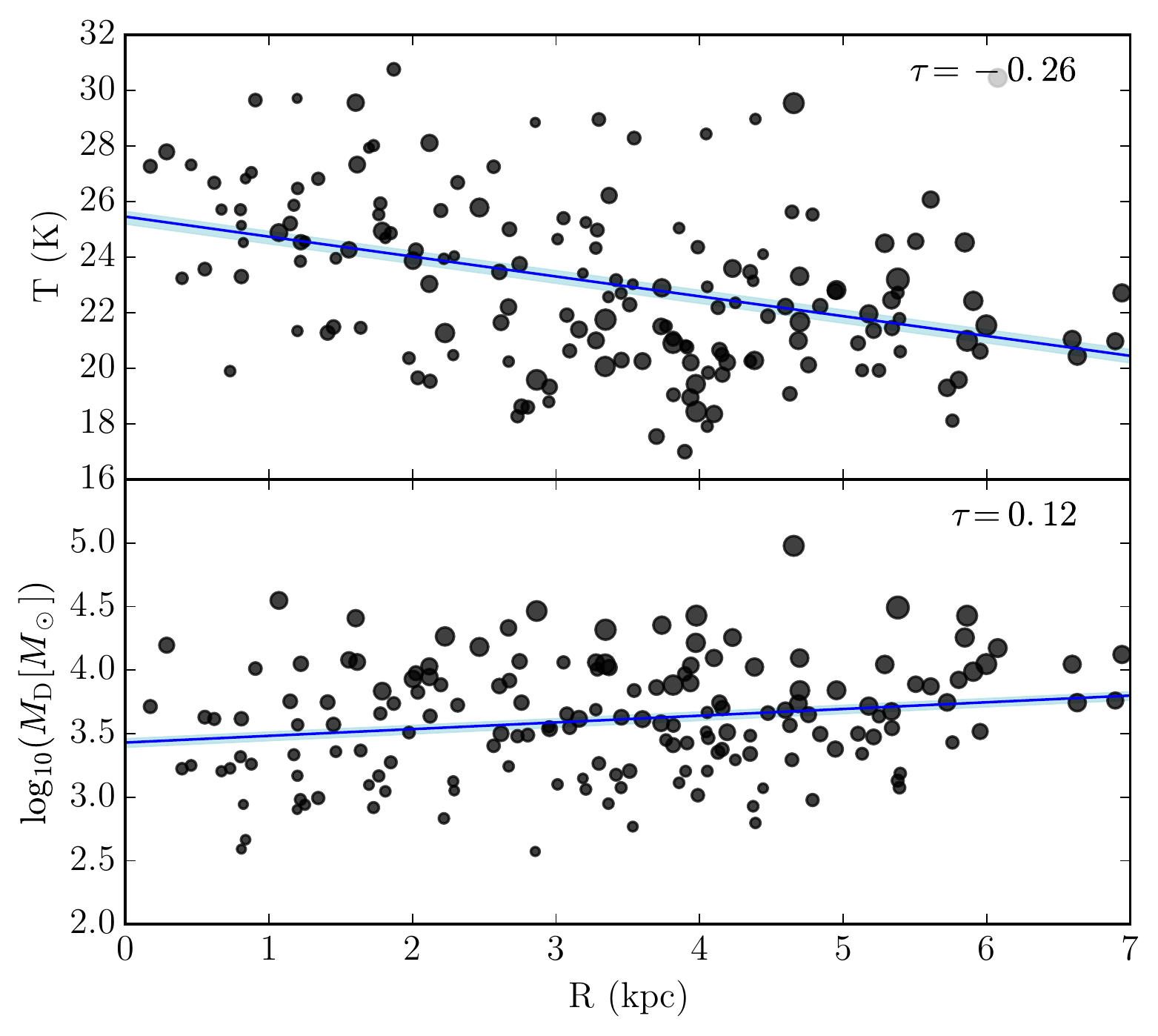}
    \caption{Radial variation in fitted dust temperature and dust mass. Point size is based on the FWHM of the cloud. In each case, the Kendall rank correlation coefficient ($\tau$) is given in the top right.}
    \label{fig:gmc_t_m_radial}
\end{figure}

We also investigated the radial variation in our two fitted MBB properties -- dust temperature and dust mass (Fig. \ref{fig:gmc_t_m_radial}. In both cases, we see that the radial correlations are weak -- in the case of dust temperature, weakly negative (i.e. dust temperatures are lower at higher galactocentric radii), and weakly positive in the case of dust mass. We find a dust temperature gradient $-0.71\pm0.01$\,K\,kpc$^{-1}$, and a dust mass gradient of $0.053\pm0.001$\,dex\,kpc$^{-1}$. The decrease in dust temperature is naturally explained by a general decrease in the strength of the interstellar radiation field (ISRF) at increasing galactocentric radius \citep{1990Rice}. This gradient is also similar to that seen by \cite{2014Tabatabaei}, when considering the global properties of M33. The invariance in dust mass is likely due to a balance of generally more compact but brighter clouds in the centre of the galaxy, whereas in the outskirts we tend to find somewhat more diffuse (but extended) sources.

\section{Comparison to CO}\label{sec:comparison_co}

\begin{figure}
	\includegraphics[width=\columnwidth]{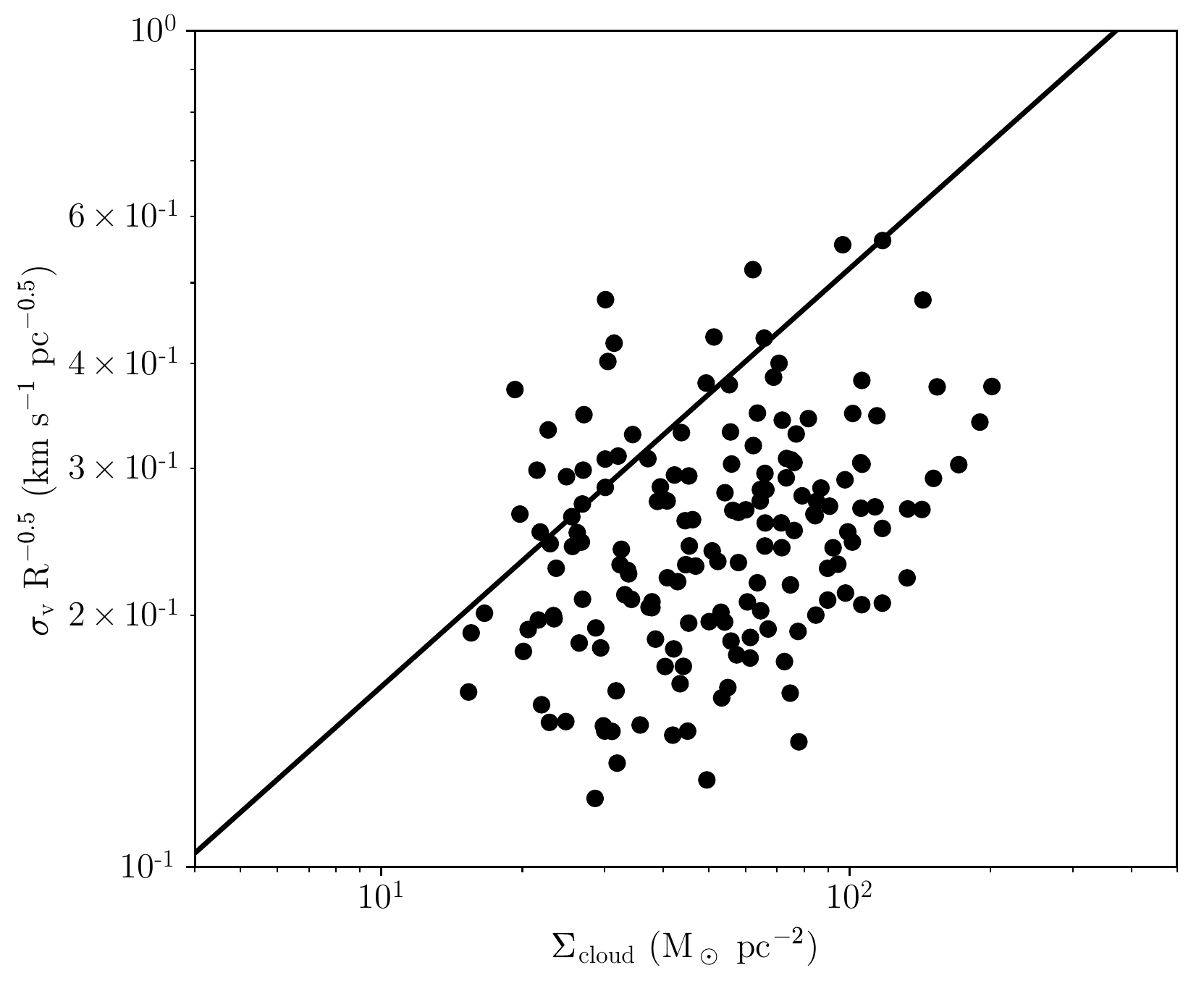}
    \caption{The ratio of velocity dispersion to the square root of the cloud radius as a function of the cloud surface density (black points). The black line shows the expected value for virialised clouds.}
    \label{fig:larsons_laws}
\end{figure}

Using kinematic CO data, we can relate the properties of these clouds to the observed scaling relations of \cite{1981Larson}. This can be neatly represented in the plane of the cloud surface density ($\Sigma_{\rm cloud}$) with the ratio of the velocity dispersion ($\sigma_{\rm v}$) to the square root of the cloud radius (R), as demonstrated in \cite{2009Heyer}. If these clouds are ideally virialised (Larson's second law), and the velocity dispersion is related to the cloud's radius as $\sigma_{\rm v} \propto R^{1/2}$ (Larson's first law), then it can be shown
\begin{equation}
\sigma_{\rm v} = \left(\frac{\pi G}{5}\right)^{1/2} \Sigma^{1/2} R^{1/2}.
\end{equation}
Larson's third law states that $\Sigma_{\rm cloud}$ is approximately equal for any cloud, so we would expect little dynamic range in this quantity. We calculate the velocity dispersion from the CO data cube of \cite{2012Gratier}, following the ``equivalent width'' as defined in \cite{2001Heyer}:
\begin{equation}
\sigma_{\rm v} = \frac{I_{\rm CO}}{\sqrt{2\pi} T_{\rm peak}},
\end{equation}
with $I_{\rm CO}$ the integrated CO in each leaf contour, and $T_{\rm peak}$ the peak line intensity. This relationship is shown in Fig. \ref{fig:larsons_laws}. We find that there is a weak correlation ($\tau$ = 0.21) between $\sigma_{\rm v}/R^{0.5}$ and $\Sigma_{\rm cloud}$ for our clouds. This is somewhat weaker than that found by \cite{2009Heyer} for a selection of clouds in the MW, but still showing a dependence of $\sigma_{\rm v}/R^{0.5}$ on the gas surface density. Given the proximity of these clouds to the line of virial equilibrium (the median deviation below this line is a factor of 1.5), we conclude that these clouds are likely ideally virialised.

\begin{figure*}
	\includegraphics[width=2\columnwidth]{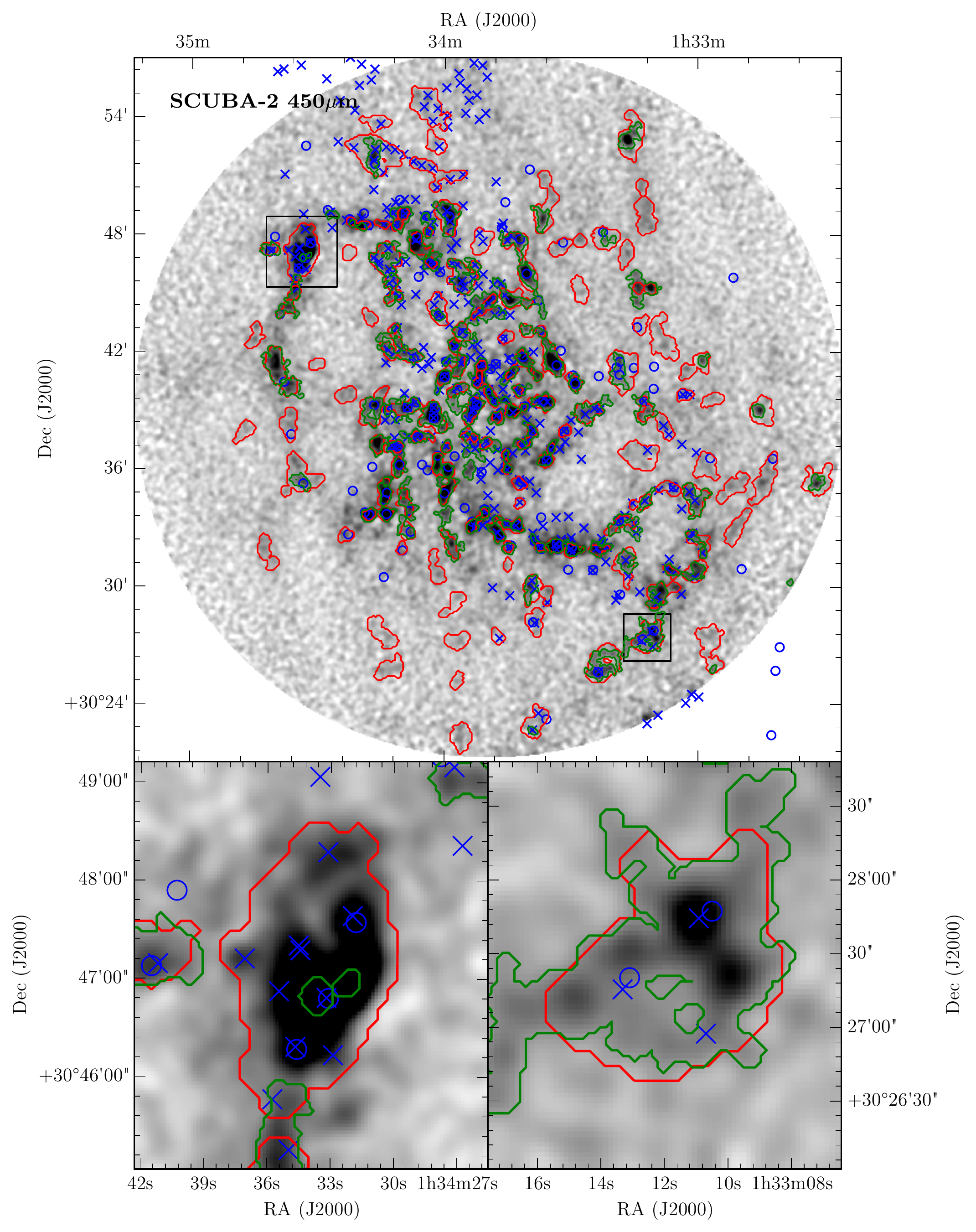}
    \caption{\textit{Top:} Comparison to earlier GMC studies of M33, overlaid on the slightly smoothed SCUBA-2 450\micron\, data. The red contours are our clouds defined from the SPIRE 250\micron \,source extraction, the green contours the SCUBA-2 450\micron\, source extraction, the blue crosses the GMCs of \citet{2012Gratier}, and the blue circles the GMCs of \citet{2003Engargiola}. \textit{Lower left:} Zoom-in of top-left rectangle (NGC 604). \textit{Lower right:} Zoom-in of lower-right rectangle. The symbols used are the same as in the top panel.}
    \label{fig:co_comparison}
\end{figure*}

Finally, we make comparison to the locations of GMCs identified with earlier CO studies. Fig. \ref{fig:co_comparison} shows the positions of our clouds against those of \cite{2010Gratier} and \cite{2003Engargiola}. These earlier studies have somewhat better resolution than we have achieved in this investiation (18\,arcsec versus 12\,arcsec), but clearly the cloud distribution is broadly similar, indicating that these particularly dusty regions are indeed associated with GMCs. However, what often appears to be a single region, even in the 450\micron\, data, is identified as several clouds in the CO data. Fig. \ref{fig:co_comparison} shows that at the 450\micron\, resolution, larger sources are beginning to break up into smaller clouds, but this is not statistically significant enough for the extraction criteria to define them as seperate leaves. Some of these sources may also be co-spatial along the line-of-sight of the galaxy, but given no kinematic information we cannot separate these as in the CO surveys.

One notable difference between our detected sources and earlier CO surveys is that the works of \cite{2003Engargiola} and \cite{2012Gratier} find a dearth of massive clouds beyond a galactocentric radius of 4kpc. However, we find a nearly flat distribution of dust mass with galactocentric radius. Given that we find our clouds to be co-spatial to these earlier works in the inner region of M33, we would expect these earlier surveys to detect these clouds. We do not believe this is a selection effect due to noise in these CO maps. \cite{2012Gratier} map an area significantly beyond 4kpc, with similar noise as in the centre of the map (see Fig. 3 of \citealt{2010Gratier}). \cite{2003Engargiola} estimate their map to be complete out to 5.2kpc and more than 50\% complete up to 8kpc, so this variation cannot simply be attributed to completeness. Results from {\it Planck} \citep{2011PlanckXIX} have shown a significant reservoir of molecular hydrogen that is not traced by CO. \cite{2017Gratier} find that this ``CO-dark'' gas forms around 50\% of the total molecular hydrogen mass of M33. The amount of CO dark gas is also expected to increase at lower metallicity (such as in the outskirts of M33), where the CO is more susceptible to photo-dissociation. Given that the dust continuum is not subject to these same caveats, the dust may offer a more representative view of the cloud population than CO surveys in these lower-metallicity environments.

\section{Conclusions}\label{sec:conclusions}

In this work, we have combined archival \textit{Herschel} FIR and sub-mm data with deep SCUBA-2 observations to probe the properties of GMCs using their dust content. Using wavelengths from 100 to 850\micron, we have probed the cold dust continuum emission of these sources. We performed source extraction using dendrograms, which found a total of 165 GMCs with sizes (FWHM) of 46-280\,pc, and a median size of 105\,pc.  By fitting a one-temperature MBB, we have calculated the dust mass and temperature for these 165 sources, and compared this to archival CO and H{\sc i} data. Using a method similar to that of \cite{2013Sandstrom}, we find a weak radial variation in the GDR of these sources, and use this GDR to calculate a total cloud mass.

These cloud masses span the range of $10^4-10^7 {\rm M}_\odot$, and the mass function can be fit with a power law slope proportional to ${\rm M}^{-2.84}$, steeper than seen in previous CO studies of M33 and the MW. Whilst we can rule out pressure as the major driver of this inefficient cloud assembly, we are unable to distinguish whether metallicity or turbulent H{\sc i} velocities contribute more to this inefficiency. The dust temperatures of these clouds range from 17-32\,K, and dust masses from 10$^2$-10$^5 \,{\rm M}_\odot$. In terms of these clouds' dust properties, we find only weak radial trends with dust mass and dust temperature.

A comparison to CO data shows an $\alpha_{\rm CO}$ factor several times higher than found in nearby galaxies. We attribute this to the subsolar metallicity of M33, where CO is likely a less suitable tracer of molecular hydrogen. We have examined these clouds in the framework of Larson's scaling relations, and we find a dependence of $\sigma_{\rm v}/R^{0.5}$ with the gas surface density, much like that of \cite{2009Heyer}. It would also appear that our clouds are ideally virialised (to within a median factor of 1.5). Finally, a comparison with earlier CO studies shows that the GMCs we are detecting and those found using CO data are generally co-spatial, but due to the limited resolution of the SPIRE 250\micron \,data that we convolve and regrid to, the crowded complexes of clouds seen in this CO data is generally confused into one large cloud in our source extraction. We also find clouds beyond 4kpc in galactocentric radius, unlike earlier CO surveys. This may be due to the CO being a poorer tracer of molecular hydrogen at these larger distances (and lower metallicities), and these clouds being dominated by CO-dark gas. 

\section*{Acknowledgements}

M.W.L.S acknowledges support from the European  Research  Council  (ERC)  Forward  Progress  7 (FP7)  project  HELP. The James Clerk Maxwell Telescope is operated by the East Asian Observatory on behalf of The National Astronomical Observatory of Japan, Academia Sinica Institute of Astronomy and Astrophysics, the Korea Astronomy and Space Science Institute, the National Astronomical Observatories of China and the Chinese Academy of Sciences (Grant No. XDB09000000), with additional funding support from the Science and Technology Facilities Council of the United Kingdom and participating universities in the United Kingdom and Canada.

This research made use of {\sc montage} (\url{http://montage.ipac.caltech.edu/ }), which is funded by the National Science Foundation under Grant Number ACI-1440620, and was previously funded by the National Aeronautics and Space Administration's Earth Science Technology Office, Computation Technologies Project, under Cooperative Agreement Number NCC5-626 between NASA and the California Institute of Technology.

This research has made use of Astropy, a community-developed core Python package for Astronomy (\url{http://www.astropy.org/}; \citealt{2013Astropy,2018Astropy}). This research has made use of NumPy (\url{http://www.numpy.org/}; \citealt{2011vanderWalt}), SciPy (\url{http://www.scipy.org/}), and MatPlotLib (\url{http://matplotlib.org/}; \citealt{2007Hunter}). This research made use of APLpy, an open-source plotting package for Python (\url{https://aplpy.github.io/}; \citealt{2012Robitaille}).



\bibliographystyle{mnras}
\bibliography{bibliography}



\appendix

\section{Source Extraction Comparison}\label{app:source_extraction_comparison}

\begin{figure*}
	\includegraphics[width=2\columnwidth]{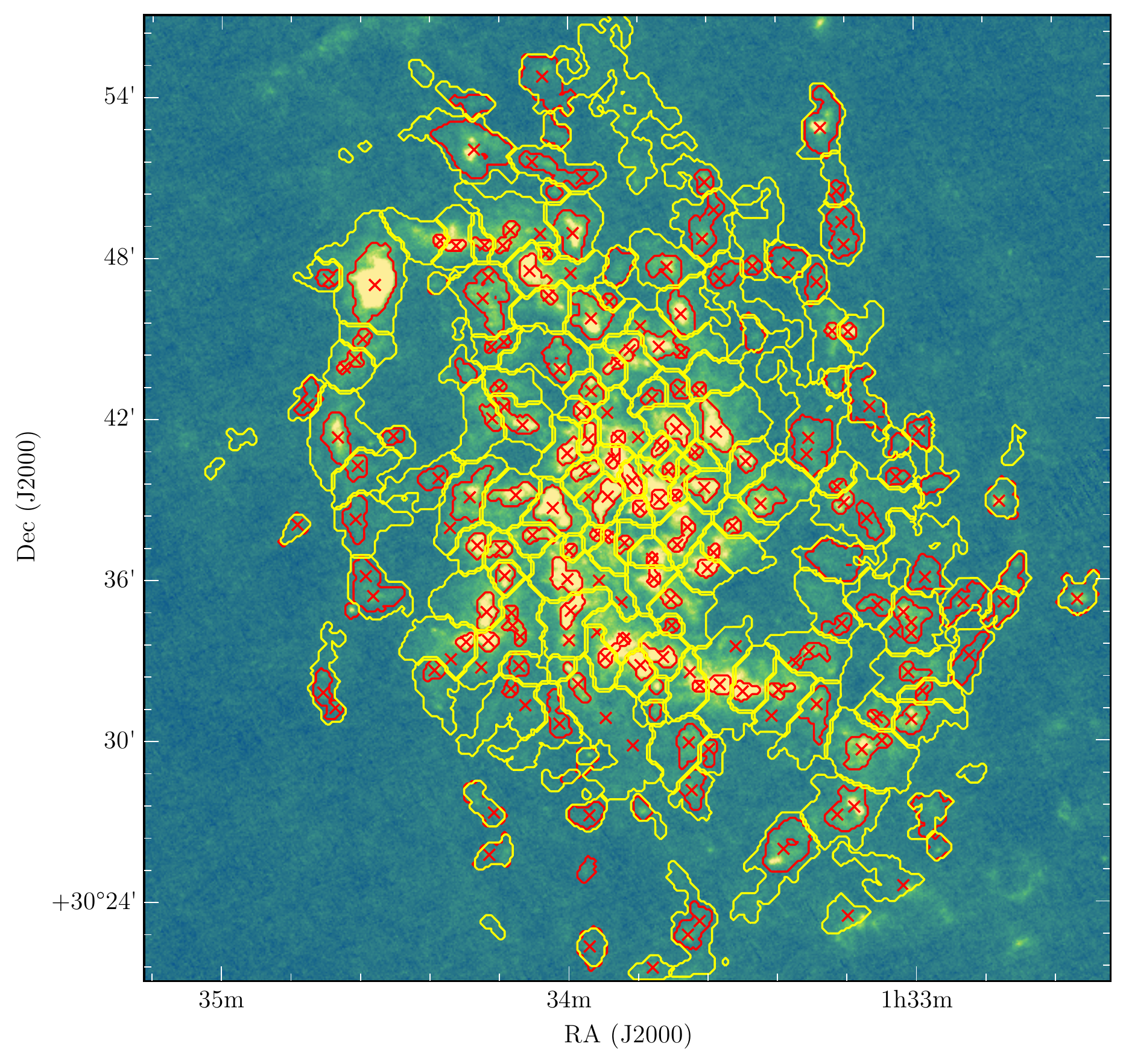}
    \caption{Comparison of 3 extraction algorithms overlaid on the PACS 160\micron\, map. Yellow contours indicate sources as detected by {\sc FellWalker}, red contours by {\sc astrodendro} and red crosses from {\sc SExtractor}.}
    \label{fig:extraction_comparison}
\end{figure*}

Here, we present a brief overview of the various source extraction algorithms used in our initial testing -- {\sc fellwalker}, {\sc SExtractor}, and {\sc astrodendro}. We have attempted to homogenise the extraction criteria to make testing these algorithms as fair as possible. In all cases, we detect objects only if they are 3$\sigma$ above the background, with an area larger than the beam. For {\sc fellwalker} and {\sc astrodendro}, we also set a minimum significance for the structure to be 3$\sigma$ (else the peaks will be merged into a single peak). For {\sc SExtractor}, we turn off the deblending threshold. The results of these various algorithms are shown in Fig. \ref{fig:extraction_comparison}. There is good correspondence between the three algorithms, and each detect a similar number of sources (169 for {\sc fellwalker}, 165 for {\sc astrodendro}, and 188 for {\sc SExtractor}). However, we can see that {\sc fellwalker} essentially partitions all of the emission in the image, leading to clearly unreasonably large structures. {\sc astrodendro}, however, finds more compact sources of emission. {\sc SExtractor} can deblend some of the most crowded regions, resolving single sources in the other algorithms into several smaller sources, but produces an ellipse of emission, rather than an exact contour. Given that {\sc SExtractor} can have overlapping ellipses (which is always the case in these crowded areas), whereas {\sc astrodendro} separates sources by default, we have opted to use {\sc astrodendro} in this work.

\section{Leaf Node Parameters}\label{app:leaf_node_parameters}
\begin{landscape}
\begin{table}
\caption{Leaf node parameters for the GMCs. Generally, errors given are 1$\sigma$ errors. However, in the case of an 3$\sigma$ upper limit, the given flux is 0 and the error given is the 3$\sigma$ upper limit.}
\label{table:leaf_node_parameters}
\begin{tabular}{ccccccccccccccc}
\hline \hline
GMC ID & R.A. (J2000) & Dec (J2000) & R (kpc) & FWHM (pc) & $S_{100}$ (Jy) & $\sigma_{100}$ & $S_{160}$ (Jy) & $\sigma_{160}$ & $S_{250}$ (Jy) & $\sigma_{250}$ & $S_{450}$ (Jy) & $\sigma_{450}$ & $S_{850}$ (Jy) & $\sigma_{850}$ \\
\hline
0 & 1 33 56.0 & 30 22 22.4 & 6.9 & 152 & 0.43 & 0.01 & 0.43 & 0.02 & 0.55 & 0.01 & 0.135 & 0.009 & 0.036 & 0.004 \\
1 & 1 33 38.5 & 30 23 04.7 & 6.6 & 172 & 0.83 & 0.03 & 1.24 & 0.03 & 0.84 & 0.01 & 0.151 & 0.008 & 0.046 & 0.002 \\
2 & 1 33 56.9 & 30 25 14.2 & 5.8 & 87 & 0.06 & 0.01 & 0.18 & 0.01 & 0.15 & 0.01 & 0.020 & 0.002 & 0.005 & 0.001 \\
3 & 1 33 22.7 & 30 26 02.6 & 5.9 & 236 & 1.97 & 0.04 & 2.83 & 0.05 & 1.97 & 0.02 & 0.495 & 0.012 & 0.142 & 0.004 \\
4 & 1 34 12.8 & 30 25 50.0 & 5.8 & 155 & 0.37 & 0.02 & 0.64 & 0.02 & 0.54 & 0.01 & 0.087 & 0.009 & 0.032 & 0.002 \\
5 & 1 32 56.3 & 30 26 02.0 & 7.1 & 105 & 0.17 & 0.01 & 0.21 & 0.01 & 0.19 & 0.01 & 0.021 & 0.005 & 0.010 & 0.002 \\
6 & 1 32 57.5 & 30 27 22.9 & 6.6 & 178 & 0.34 & 0.02 & 0.54 & 0.02 & 0.38 & 0.01 & 0.077 & 0.009 & 0.028 & 0.002 \\
7 & 1 33 11.6 & 30 27 27.5 & 5.8 & 195 & 3.42 & 0.04 & 3.40 & 0.05 & 2.11 & 0.02 & 0.444 & 0.010 & 0.130 & 0.003 \\
8 & 1 33 56.8 & 30 27 17.2 & 4.9 & 134 & 0.25 & 0.02 & 0.38 & 0.02 & 0.28 & 0.01 & 0.038 & 0.003 & 0 & 0.004 \\
9 & 1 34 14.5 & 30 27 39.2 & 5.2 & 176 & 0.47 & 0.03 & 0.69 & 0.02 & 0.52 & 0.01 & 0.076 & 0.007 & 0.009 & 0.001 \\
10 & 1 33 47.3 & 30 27 28.6 & 4.8 & 85 & 0.22 & 0.01 & 0.22 & 0.02 & 0.14 & 0.01 & 0 & 0.012 & 0.004 & 0.001 \\
11 & 1 33 38.6 & 30 28 14.9 & 4.6 & 140 & 0.51 & 0.02 & 0.68 & 0.02 & 0.43 & 0.01 & 0.096 & 0.006 & 0.013 & 0.001 \\
12 & 1 33 57.5 & 30 29 12.3 & 4.2 & 144 & 0.18 & 0.02 & 0.30 & 0.02 & 0.22 & 0.01 & 0.040 & 0.005 & 0.010 & 0.002 \\
13 & 1 33 39.2 & 30 29 56.4 & 3.9 & 146 & 0.61 & 0.03 & 0.96 & 0.03 & 0.73 & 0.01 & 0.152 & 0.006 & 0.044 & 0.001 \\
14 & 1 33 09.0 & 30 29 42.3 & 5.3 & 180 & 2.07 & 0.07 & 2.09 & 0.06 & 1.34 & 0.03 & 0.211 & 0.012 & 0.050 & 0.004 \\
15 & 1 33 35.8 & 30 29 39.5 & 4.1 & 99 & 0.24 & 0.02 & 0.25 & 0.02 & 0.23 & 0.01 & 0.039 & 0.004 & 0.007 & 0.001 \\
16 & 1 33 17.2 & 30 31 13.0 & 4.4 & 177 & 0.61 & 0.03 & 1.02 & 0.02 & 0.72 & 0.02 & 0.142 & 0.007 & 0.029 & 0.001 \\
17 & 1 34 01.8 & 30 30 53.6 & 3.6 & 150 & 0.23 & 0.01 & 0.44 & 0.02 & 0.30 & 0.01 & 0.041 & 0.005 & 0.010 & 0.001 \\
18 & 1 33 00.7 & 30 30 49.4 & 5.5 & 139 & 1.49 & 0.03 & 1.49 & 0.04 & 0.90 & 0.02 & 0.201 & 0.008 & 0.043 & 0.002 \\
19 & 1 33 06.5 & 30 30 49.2 & 5.1 & 81 & 0.12 & 0.02 & 0.16 & 0.02 & 0.15 & 0.01 & 0.051 & 0.004 & 0.006 & 0.001 \\
20 & 1 33 45.0 & 30 31 09.9 & 3.4 & 64 & 0.10 & 0.01 & 0.12 & 0.02 & 0.10 & 0.01 & 0.011 & 0.003 & 0.003 & 0.000 \\
21 & 1 34 40.4 & 30 31 13.2 & 5.4 & 79 & 0.10 & 0.01 & 0.18 & 0.01 & 0.12 & 0.01 & 0.015 & 0.003 & 0 & 0.003 \\
22 & 1 34 42.4 & 30 32 01.4 & 5.3 & 118 & 0.30 & 0.02 & 0.35 & 0.02 & 0.32 & 0.01 & 0.046 & 0.005 & 0.017 & 0.001 \\
23 & 1 33 58.2 & 30 32 04.1 & 3.1 & 103 & 0.41 & 0.04 & 0.55 & 0.04 & 0.42 & 0.01 & 0.077 & 0.007 & 0.011 & 0.001 \\
24 & 1 33 29.7 & 30 31 52.5 & 3.5 & 94 & 2.87 & 0.08 & 2.43 & 0.10 & 1.13 & 0.05 & 0.176 & 0.013 & 0.038 & 0.003 \\
25 & 1 32 58.7 & 30 31 52.1 & 5.4 & 76 & 0.10 & 0.01 & 0.12 & 0.02 & 0.13 & 0.01 & 0.024 & 0.003 & 0.004 & 0.001 \\
26 & 1 33 23.7 & 30 31 55.4 & 3.8 & 95 & 0.14 & 0.03 & 0.25 & 0.03 & 0.21 & 0.02 & 0.056 & 0.005 & 0.010 & 0.001 \\
27 & 1 34 10.0 & 30 31 58.9 & 3.5 & 70 & 0.14 & 0.01 & 0.17 & 0.01 & 0.12 & 0.01 & 0.019 & 0.004 & 0.005 & 0.001 \\
28 & 1 33 44.9 & 30 32 04.9 & 3.0 & 62 & 0.24 & 0.02 & 0.25 & 0.03 & 0.19 & 0.01 & 0.018 & 0.005 & 0.004 & 0.001 \\
29 & 1 33 33.9 & 30 32 07.7 & 3.3 & 97 & 2.13 & 0.06 & 2.25 & 0.08 & 1.24 & 0.03 & 0.202 & 0.007 & 0.040 & 0.002 \\
30 & 1 33 37.4 & 30 32 04.4 & 3.2 & 53 & 0.20 & 0.03 & 0.23 & 0.04 & 0.17 & 0.02 & 0.029 & 0.004 & 0.005 & 0.001 \\
31 & 1 32 51.0 & 30 33 02.6 & 5.7 & 160 & 0.26 & 0.02 & 0.29 & 0.02 & 0.37 & 0.01 & 0.108 & 0.007 & 0.020 & 0.002 \\
32 & 1 33 00.7 & 30 32 33.0 & 5.1 & 112 & 0.22 & 0.01 & 0.34 & 0.02 & 0.26 & 0.01 & 0.033 & 0.006 & 0.015 & 0.001 \\
33 & 1 34 23.2 & 30 32 39.5 & 3.9 & 93 & 0.17 & 0.03 & 0.27 & 0.02 & 0.21 & 0.01 & 0.034 & 0.004 & 0.010 & 0.001 \\
34 & 1 33 47.6 & 30 32 51.4 & 2.7 & 109 & 1.75 & 0.08 & 1.74 & 0.09 & 1.03 & 0.03 & 0.168 & 0.006 & 0.039 & 0.002 \\
35 & 1 34 08.6 & 30 32 50.7 & 3.1 & 100 & 0.23 & 0.01 & 0.33 & 0.03 & 0.27 & 0.01 & 0.069 & 0.006 & 0.008 & 0.001 \\
36 & 1 33 59.9 & 30 32 43.4 & 2.9 & 47 & 0.17 & 0.01 & 0.16 & 0.02 & 0.09 & 0.01 & 0 & 0.008 & 0.003 & 0.001 \\
37 & 1 33 18.6 & 30 33 13.6 & 3.7 & 144 & 0.33 & 0.02 & 0.46 & 0.02 & 0.33 & 0.02 & 0.061 & 0.006 & 0.011 & 0.002 \\
38 & 1 33 53.3 & 30 33 12.5 & 2.6 & 90 & 0.86 & 0.06 & 0.76 & 0.08 & 0.43 & 0.04 & 0.059 & 0.008 & 0.013 & 0.002 \\
39 & 1 33 43.9 & 30 33 10.5 & 2.6 & 123 & 1.10 & 0.06 & 1.19 & 0.07 & 0.82 & 0.03 & 0.136 & 0.008 & 0.029 & 0.002 \\
40 & 1 34 17.7 & 30 33 43.2 & 3.3 & 75 & 0.90 & 0.05 & 0.90 & 0.06 & 0.56 & 0.03 & 0.115 & 0.006 & 0.021 & 0.001 \\
41 & 1 34 13.9 & 30 33 45.1 & 3.1 & 86 & 2.67 & 0.07 & 2.53 & 0.08 & 1.48 & 0.03 & 0.284 & 0.011 & 0.048 & 0.002 \\
42 & 1 34 09.2 & 30 34 21.1 & 2.6 & 130 & 0.24 & 0.04 & 0.42 & 0.04 & 0.29 & 0.02 & 0.077 & 0.007 & 0 & 0.006 \\
43 & 1 33 50.5 & 30 33 47.6 & 2.3 & 52 & 0.19 & 0.03 & 0.20 & 0.03 & 0.15 & 0.02 & 0.027 & 0.004 & 0.004 & 0.001 \\
44 & 1 33 01.0 & 30 34 33.2 & 4.7 & 180 & 1.76 & 0.03 & 2.12 & 0.04 & 1.28 & 0.02 & 0.222 & 0.008 & 0.040 & 0.002 \\
\end{tabular}
\end{table}
\end{landscape}
\begin{landscape}
\begin{table}
\contcaption{Leaf node parameters for the GMCs.}
\label{table:leaf_node_parameters_b}
\begin{tabular}{ccccccccccccccc}
\hline \hline
GMC ID & R.A. (J2000) & Dec (J2000) & R (kpc) & FWHM (pc) & $S_{100}$ (Jy) & $\sigma_{100}$ & $S_{160}$ (Jy) & $\sigma_{160}$ & $S_{250}$ (Jy) & $\sigma_{250}$ & $S_{450}$ (Jy) & $\sigma_{450}$ & $S_{850}$ (Jy) & $\sigma_{850}$ \\
\hline
45 & 1 33 42.2 & 30 34 18.3 & 2.2 & 63 & 0.10 & 0.01 & 0.12 & 0.01 & 0.11 & 0.01 & 0.011 & 0.003 & 0 & 0.002 \\
46 & 1 33 13.3 & 30 34 20.5 & 3.8 & 114 & 0.19 & 0.01 & 0.28 & 0.02 & 0.21 & 0.01 & 0.076 & 0.006 & 0 & 0.005 \\
47 & 1 34 14.2 & 30 34 51.3 & 2.7 & 123 & 1.83 & 0.04 & 2.04 & 0.06 & 1.31 & 0.03 & 0.207 & 0.010 & 0.041 & 0.002 \\
48 & 1 33 59.5 & 30 34 53.3 & 2.0 & 114 & 1.67 & 0.09 & 1.80 & 0.08 & 1.08 & 0.04 & 0.219 & 0.010 & 0.032 & 0.002 \\
49 & 1 32 44.2 & 30 35 31.5 & 5.9 & 199 & 1.08 & 0.03 & 1.18 & 0.02 & 0.96 & 0.01 & 0.226 & 0.009 & 0.053 & 0.003 \\
50 & 1 32 51.8 & 30 35 08.2 & 5.3 & 165 & 0.54 & 0.02 & 0.60 & 0.02 & 0.46 & 0.01 & 0.091 & 0.007 & 0.025 & 0.001 \\
51 & 1 34 37.4 & 30 34 53.9 & 4.4 & 62 & 0.29 & 0.01 & 0.23 & 0.01 & 0.12 & 0.01 & 0.014 & 0.003 & 0.003 & 0.000 \\
52 & 1 33 06.9 & 30 35 04.0 & 4.2 & 119 & 0.25 & 0.01 & 0.34 & 0.02 & 0.37 & 0.01 & 0.058 & 0.006 & 0.020 & 0.001 \\
53 & 1 32 31.8 & 30 35 21.8 & 6.9 & 173 & 1.58 & 0.03 & 1.87 & 0.02 & 1.28 & 0.01 & 0.248 & 0.009 & 0.053 & 0.002 \\
54 & 1 34 33.8 & 30 35 42.1 & 4.0 & 237 & 0.78 & 0.03 & 1.64 & 0.03 & 1.36 & 0.01 & 0.302 & 0.007 & 0.065 & 0.002 \\
55 & 1 33 42.3 & 30 35 18.5 & 1.8 & 83 & 0.37 & 0.03 & 0.41 & 0.03 & 0.26 & 0.01 & 0.031 & 0.006 & 0.007 & 0.001 \\
56 & 1 34 00.3 & 30 36 06.4 & 1.6 & 146 & 3.99 & 0.11 & 3.43 & 0.09 & 1.77 & 0.03 & 0.296 & 0.009 & 0.052 & 0.001 \\
57 & 1 32 58.2 & 30 36 11.6 & 4.7 & 169 & 0.37 & 0.02 & 0.68 & 0.02 & 0.43 & 0.01 & 0.079 & 0.006 & 0.008 & 0.002 \\
58 & 1 34 11.0 & 30 36 10.2 & 2.2 & 99 & 1.86 & 0.04 & 1.77 & 0.04 & 1.03 & 0.02 & 0.152 & 0.009 & 0.034 & 0.002 \\
59 & 1 33 45.2 & 30 36 06.3 & 1.5 & 66 & 0.34 & 0.07 & 0.44 & 0.05 & 0.27 & 0.02 & 0.047 & 0.005 & 0.008 & 0.001 \\
60 & 1 33 14.0 & 30 36 44.3 & 3.3 & 220 & 0.61 & 0.03 & 0.83 & 0.03 & 0.81 & 0.01 & 0.179 & 0.009 & 0.040 & 0.002 \\
61 & 1 33 36.0 & 30 36 28.0 & 1.8 & 88 & 1.19 & 0.08 & 1.18 & 0.07 & 0.60 & 0.04 & 0.106 & 0.007 & 0.025 & 0.001 \\
62 & 1 33 45.6 & 30 36 48.4 & 1.2 & 46 & 0.40 & 0.06 & 0.32 & 0.06 & 0.19 & 0.02 & 0.019 & 0.005 & 0.005 & 0.001 \\
63 & 1 34 15.8 & 30 37 19.0 & 2.3 & 97 & 1.62 & 0.05 & 1.46 & 0.06 & 0.74 & 0.03 & 0.167 & 0.006 & 0.024 & 0.001 \\
64 & 1 33 59.7 & 30 37 06.6 & 1.3 & 60 & 0.16 & 0.02 & 0.18 & 0.02 & 0.14 & 0.01 & 0.027 & 0.003 & 0.002 & 0.000 \\
65 & 1 33 34.8 & 30 37 04.2 & 1.7 & 54 & 0.48 & 0.03 & 0.37 & 0.03 & 0.22 & 0.02 & 0.033 & 0.004 & 0.006 & 0.001 \\
66 & 1 34 11.9 & 30 37 10.2 & 2.0 & 97 & 0.31 & 0.04 & 0.53 & 0.04 & 0.43 & 0.02 & 0.084 & 0.008 & 0.023 & 0.001 \\
67 & 1 33 50.3 & 30 37 23.4 & 0.9 & 70 & 0.57 & 0.03 & 0.51 & 0.03 & 0.34 & 0.02 & 0.037 & 0.004 & 0.006 & 0.001 \\
68 & 1 33 41.1 & 30 37 21.8 & 1.2 & 72 & 0.15 & 0.01 & 0.21 & 0.02 & 0.16 & 0.02 & 0.014 & 0.004 & 0.003 & 0.001 \\
69 & 1 34 47.1 & 30 37 58.7 & 4.8 & 121 & 0.32 & 0.01 & 0.45 & 0.02 & 0.31 & 0.01 & 0.048 & 0.004 & 0.006 & 0.001 \\
70 & 1 34 36.7 & 30 38 19.8 & 3.9 & 155 & 0.28 & 0.02 & 0.51 & 0.02 & 0.48 & 0.01 & 0.076 & 0.006 & 0.028 & 0.002 \\
71 & 1 34 05.3 & 30 37 42.1 & 1.4 & 107 & 0.30 & 0.04 & 0.46 & 0.05 & 0.32 & 0.02 & 0.074 & 0.008 & 0.009 & 0.001 \\
72 & 1 33 53.0 & 30 37 38.5 & 0.8 & 46 & 0.08 & 0.01 & 0.12 & 0.02 & 0.08 & 0.01 & 0.013 & 0.003 & 0 & 0.001 \\
73 & 1 33 55.0 & 30 37 42.7 & 0.8 & 52 & 0.13 & 0.02 & 0.18 & 0.02 & 0.11 & 0.01 & 0 & 0.009 & 0.003 & 0.001 \\
74 & 1 33 08.0 & 30 38 12.4 & 3.7 & 126 & 0.13 & 0.02 & 0.35 & 0.03 & 0.32 & 0.01 & 0.077 & 0.005 & 0.016 & 0.001 \\
75 & 1 33 39.4 & 30 37 59.4 & 1.2 & 71 & 0.54 & 0.05 & 0.53 & 0.04 & 0.33 & 0.03 & 0.045 & 0.005 & 0.010 & 0.001 \\
76 & 1 33 31.6 & 30 38 00.5 & 1.8 & 73 & 0.32 & 0.02 & 0.44 & 0.03 & 0.28 & 0.02 & 0.024 & 0.004 & 0.003 & 0.001 \\
77 & 1 34 02.7 & 30 38 42.1 & 1.1 & 166 & 7.36 & 0.10 & 7.82 & 0.10 & 4.28 & 0.03 & 0.659 & 0.012 & 0.139 & 0.003 \\
78 & 1 33 53.2 & 30 39 05.7 & 0.3 & 132 & 5.89 & 0.13 & 5.15 & 0.09 & 2.45 & 0.03 & 0.425 & 0.011 & 0.062 & 0.002 \\
79 & 1 34 16.9 & 30 39 09.4 & 2.2 & 199 & 1.45 & 0.04 & 1.98 & 0.06 & 1.50 & 0.02 & 0.278 & 0.011 & 0.062 & 0.002 \\
80 & 1 32 45.1 & 30 38 58.0 & 5.6 & 154 & 1.97 & 0.02 & 1.70 & 0.02 & 1.04 & 0.01 & 0.229 & 0.006 & 0.054 & 0.002 \\
81 & 1 33 47.6 & 30 38 41.8 & 0.5 & 65 & 0.60 & 0.06 & 0.57 & 0.07 & 0.29 & 0.03 & 0.043 & 0.005 & 0.009 & 0.001 \\
82 & 1 33 26.2 & 30 38 56.7 & 2.1 & 153 & 1.41 & 0.04 & 1.66 & 0.04 & 1.05 & 0.02 & 0.205 & 0.008 & 0.045 & 0.002 \\
83 & 1 33 12.3 & 30 38 53.3 & 3.3 & 92 & 0.82 & 0.04 & 0.67 & 0.03 & 0.39 & 0.02 & 0.031 & 0.006 & 0.011 & 0.001 \\
84 & 1 33 44.2 & 30 39 01.0 & 0.6 & 85 & 1.21 & 0.07 & 1.17 & 0.07 & 0.65 & 0.03 & 0.091 & 0.009 & 0.013 & 0.001 \\
85 & 1 34 09.1 & 30 39 11.4 & 1.6 & 140 & 2.14 & 0.06 & 2.36 & 0.06 & 1.39 & 0.03 & 0.227 & 0.009 & 0.042 & 0.001 \\
86 & 1 33 36.6 & 30 39 27.2 & 1.2 & 121 & 2.15 & 0.07 & 2.23 & 0.07 & 1.32 & 0.03 & 0.260 & 0.010 & 0.041 & 0.003 \\
87 & 1 33 41.5 & 30 39 10.5 & 0.8 & 46 & 0.16 & 0.05 & 0.17 & 0.04 & 0.13 & 0.02 & 0 & 0.017 & 0.004 & 0.001 \\
88 & 1 33 49.0 & 30 39 45.6 & 0.2 & 98 & 1.72 & 0.08 & 1.52 & 0.08 & 0.85 & 0.04 & 0.113 & 0.007 & 0.021 & 0.002 \\
89 & 1 32 56.6 & 30 39 35.0 & 4.6 & 108 & 0.13 & 0.01 & 0.30 & 0.01 & 0.21 & 0.01 & 0.033 & 0.005 & 0.011 & 0.001 \\
\end{tabular}
\end{table}
\end{landscape}
\begin{landscape}
\begin{table}
\contcaption{Leaf node parameters for the GMCs.}
\label{table:leaf_node_parameters_c}
\begin{tabular}{ccccccccccccccc}
\hline \hline
GMC ID & R.A. (J2000) & Dec (J2000) & R (kpc) & FWHM (pc) & $S_{100}$ (Jy) & $\sigma_{100}$ & $S_{160}$ (Jy) & $\sigma_{160}$ & $S_{250}$ (Jy) & $\sigma_{250}$ & $S_{450}$ (Jy) & $\sigma_{450}$ & $S_{850}$ (Jy) & $\sigma_{850}$ \\
\hline
90 & 1 33 13.3 & 30 39 30.4 & 3.2 & 66 & 0.25 & 0.02 & 0.25 & 0.03 & 0.17 & 0.01 & 0.029 & 0.004 & 0 & 0.004 \\
91 & 1 34 23.3 & 30 39 48.9 & 2.8 & 120 & 0.18 & 0.02 & 0.33 & 0.02 & 0.30 & 0.01 & 0.073 & 0.006 & 0.014 & 0.002 \\
92 & 1 33 03.3 & 30 39 50.5 & 4.1 & 88 & 0.13 & 0.01 & 0.29 & 0.01 & 0.21 & 0.01 & 0.039 & 0.004 & 0.005 & 0.001 \\
93 & 1 34 36.5 & 30 40 14.3 & 3.9 & 108 & 0.15 & 0.02 & 0.36 & 0.02 & 0.37 & 0.01 & 0.118 & 0.005 & 0.018 & 0.001 \\
94 & 1 33 57.0 & 30 40 06.3 & 0.6 & 95 & 0.62 & 0.03 & 0.77 & 0.04 & 0.48 & 0.02 & 0.067 & 0.005 & 0.014 & 0.001 \\
95 & 1 33 42.6 & 30 40 02.4 & 0.7 & 63 & 0.08 & 0.03 & 0.14 & 0.02 & 0.12 & 0.01 & 0.032 & 0.005 & 0.004 & 0.001 \\
96 & 1 33 18.0 & 30 41 05.2 & 2.9 & 228 & 1.32 & 0.03 & 2.23 & 0.05 & 1.81 & 0.02 & 0.372 & 0.009 & 0.118 & 0.002 \\
97 & 1 33 29.3 & 30 40 27.2 & 1.9 & 93 & 3.33 & 0.07 & 2.34 & 0.06 & 1.11 & 0.02 & 0.172 & 0.007 & 0.028 & 0.001 \\
98 & 1 33 52.1 & 30 40 35.2 & 0.4 & 76 & 0.24 & 0.05 & 0.29 & 0.05 & 0.16 & 0.02 & 0.043 & 0.007 & 0.007 & 0.002 \\
99 & 1 34 39.9 & 30 41 19.9 & 4.2 & 167 & 2.73 & 0.07 & 2.88 & 0.07 & 1.92 & 0.02 & 0.425 & 0.009 & 0.097 & 0.003 \\
100 & 1 34 00.1 & 30 40 46.0 & 0.9 & 92 & 5.30 & 0.07 & 4.10 & 0.07 & 1.89 & 0.04 & 0.269 & 0.007 & 0.052 & 0.001 \\
101 & 1 33 38.0 & 30 40 47.7 & 1.2 & 59 & 0.12 & 0.02 & 0.17 & 0.01 & 0.13 & 0.01 & 0.030 & 0.004 & 0.004 & 0.001 \\
102 & 1 33 44.0 & 30 40 59.5 & 0.8 & 71 & 0.52 & 0.03 & 0.48 & 0.04 & 0.31 & 0.03 & 0.061 & 0.005 & 0.007 & 0.001 \\
103 & 1 33 02.6 & 30 41 02.4 & 4.2 & 101 & 0.18 & 0.02 & 0.18 & 0.01 & 0.18 & 0.01 & 0.054 & 0.005 & 0.010 & 0.001 \\
104 & 1 33 56.6 & 30 41 15.3 & 0.8 & 105 & 0.53 & 0.06 & 0.68 & 0.06 & 0.49 & 0.02 & 0.065 & 0.007 & 0.012 & 0.001 \\
105 & 1 33 34.4 & 30 41 33.7 & 1.6 & 156 & 13.27 & 0.17 & 10.37 & 0.09 & 4.52 & 0.03 & 0.771 & 0.010 & 0.141 & 0.002 \\
106 & 1 32 59.2 & 30 41 31.6 & 4.5 & 109 & 0.43 & 0.01 & 0.53 & 0.02 & 0.42 & 0.01 & 0.091 & 0.004 & 0.013 & 0.001 \\
107 & 1 33 41.2 & 30 41 37.0 & 1.1 & 109 & 1.24 & 0.04 & 1.34 & 0.06 & 0.77 & 0.03 & 0.106 & 0.007 & 0.017 & 0.002 \\
108 & 1 34 30.2 & 30 41 22.7 & 3.4 & 81 & 0.20 & 0.01 & 0.32 & 0.01 & 0.22 & 0.01 & 0 & 0.011 & 0.004 & 0.000 \\
109 & 1 33 51.3 & 30 41 18.9 & 0.7 & 59 & 0.38 & 0.05 & 0.39 & 0.04 & 0.24 & 0.02 & 0.038 & 0.005 & 0.006 & 0.001 \\
110 & 1 34 08.8 & 30 41 59.2 & 1.8 & 165 & 1.40 & 0.06 & 1.51 & 0.06 & 0.89 & 0.02 & 0.122 & 0.009 & 0.022 & 0.002 \\
111 & 1 34 13.1 & 30 42 03.0 & 2.1 & 100 & 0.19 & 0.02 & 0.35 & 0.03 & 0.28 & 0.01 & 0.053 & 0.005 & 0.013 & 0.001 \\
112 & 1 33 08.5 & 30 42 39.1 & 3.8 & 221 & 0.58 & 0.03 & 0.60 & 0.03 & 0.75 & 0.01 & 0.139 & 0.007 & 0.037 & 0.002 \\
113 & 1 33 57.7 & 30 42 17.2 & 1.2 & 75 & 1.06 & 0.04 & 0.99 & 0.05 & 0.55 & 0.02 & 0.081 & 0.003 & 0.013 & 0.001 \\
114 & 1 34 44.8 & 30 42 47.6 & 4.8 & 131 & 0.24 & 0.02 & 0.41 & 0.02 & 0.30 & 0.01 & 0.071 & 0.005 & 0.011 & 0.001 \\
115 & 1 33 45.5 & 30 42 46.8 & 1.3 & 86 & 0.27 & 0.03 & 0.33 & 0.03 & 0.22 & 0.01 & 0.012 & 0.004 & 0.004 & 0.000 \\
116 & 1 33 55.9 & 30 43 01.5 & 1.4 & 113 & 0.44 & 0.04 & 0.58 & 0.03 & 0.49 & 0.02 & 0.075 & 0.007 & 0.019 & 0.002 \\
117 & 1 33 40.7 & 30 43 05.8 & 1.6 & 81 & 0.18 & 0.02 & 0.34 & 0.03 & 0.19 & 0.01 & 0.036 & 0.003 & 0.007 & 0.001 \\
118 & 1 33 37.2 & 30 43 05.6 & 1.8 & 62 & 0.21 & 0.03 & 0.21 & 0.02 & 0.16 & 0.01 & 0.013 & 0.004 & 0.007 & 0.000 \\
119 & 1 33 13.7 & 30 43 23.4 & 3.5 & 106 & 0.25 & 0.03 & 0.19 & 0.01 & 0.15 & 0.00 & 0.035 & 0.003 & 0.011 & 0.001 \\
120 & 1 34 12.0 & 30 43 11.9 & 2.3 & 60 & 0.08 & 0.01 & 0.13 & 0.01 & 0.11 & 0.01 & 0.017 & 0.004 & 0.005 & 0.000 \\
121 & 1 34 17.6 & 30 43 46.2 & 2.8 & 95 & 0.09 & 0.01 & 0.20 & 0.01 & 0.17 & 0.01 & 0.039 & 0.003 & 0.006 & 0.001 \\
122 & 1 34 01.6 & 30 44 08.9 & 2.0 & 160 & 1.39 & 0.04 & 1.53 & 0.05 & 0.92 & 0.02 & 0.164 & 0.008 & 0.043 & 0.002 \\
123 & 1 34 37.4 & 30 44 11.1 & 4.4 & 111 & 0.33 & 0.03 & 0.39 & 0.03 & 0.22 & 0.02 & 0.060 & 0.006 & 0.009 & 0.002 \\
124 & 1 33 52.0 & 30 44 00.2 & 1.7 & 70 & 0.31 & 0.03 & 0.30 & 0.04 & 0.19 & 0.02 & 0.014 & 0.004 & 0.007 & 0.001 \\
125 & 1 33 49.4 & 30 44 33.4 & 2.0 & 79 & 0.18 & 0.04 & 0.30 & 0.04 & 0.25 & 0.02 & 0.052 & 0.005 & 0.009 & 0.001 \\
126 & 1 33 43.8 & 30 44 42.6 & 2.1 & 155 & 3.53 & 0.05 & 2.98 & 0.05 & 1.41 & 0.02 & 0.246 & 0.010 & 0.039 & 0.002 \\
127 & 1 34 12.1 & 30 44 49.7 & 2.7 & 88 & 0.07 & 0.01 & 0.22 & 0.02 & 0.15 & 0.01 & 0.031 & 0.005 & 0.006 & 0.001 \\
128 & 1 33 27.9 & 30 45 10.1 & 3.0 & 127 & 0.09 & 0.02 & 0.29 & 0.01 & 0.26 & 0.01 & 0.022 & 0.004 & 0.010 & 0.001 \\
129 & 1 34 35.5 & 30 45 00.2 & 4.4 & 79 & 0.18 & 0.01 & 0.27 & 0.02 & 0.22 & 0.02 & 0.050 & 0.007 & 0.010 & 0.001 \\
130 & 1 33 56.0 & 30 45 46.9 & 2.5 & 190 & 3.81 & 0.05 & 3.71 & 0.06 & 2.04 & 0.03 & 0.301 & 0.009 & 0.065 & 0.004 \\
131 & 1 33 14.2 & 30 45 17.2 & 3.9 & 64 & 0.28 & 0.02 & 0.30 & 0.03 & 0.17 & 0.02 & 0.027 & 0.004 & 0.007 & 0.001 \\
132 & 1 33 11.4 & 30 45 15.1 & 4.0 & 67 & 1.35 & 0.07 & 1.09 & 0.06 & 0.58 & 0.03 & 0.082 & 0.006 & 0.014 & 0.002 \\
133 & 1 34 13.7 & 30 46 31.6 & 3.3 & 243 & 1.90 & 0.08 & 2.62 & 0.07 & 1.72 & 0.03 & 0.341 & 0.014 & 0.072 & 0.004 \\
134 & 1 33 40.4 & 30 45 55.4 & 2.7 & 142 & 2.24 & 0.05 & 2.83 & 0.04 & 1.93 & 0.02 & 0.389 & 0.007 & 0.079 & 0.001 \\
\end{tabular}
\end{table}
\end{landscape}
\begin{landscape}
\begin{table}
\contcaption{Leaf node parameters for the GMCs.}
\label{table:leaf_node_parameters_d}
\begin{tabular}{ccccccccccccccc}
\hline \hline
GMC ID & R.A. (J2000) & Dec (J2000) & R (kpc) & FWHM (pc) & $S_{100}$ (Jy) & $\sigma_{100}$ & $S_{160}$ (Jy) & $\sigma_{160}$ & $S_{250}$ (Jy) & $\sigma_{250}$ & $S_{450}$ (Jy) & $\sigma_{450}$ & $S_{850}$ (Jy) & $\sigma_{850}$ \\
\hline
135 & 1 34 33.5 & 30 47 01.8 & 4.7 & 235 & 49.27 & 0.17 & 38.43 & 0.16 & 16.74 & 0.05 & 2.754 & 0.025 & 0.604 & 0.006 \\
136 & 1 33 52.8 & 30 46 23.4 & 2.7 & 63 & 0.11 & 0.01 & 0.15 & 0.01 & 0.13 & 0.01 & 0.027 & 0.003 & 0.007 & 0.001 \\
137 & 1 33 17.0 & 30 47 03.2 & 4.1 & 124 & 0.36 & 0.01 & 0.54 & 0.02 & 0.44 & 0.01 & 0.068 & 0.005 & 0.016 & 0.001 \\
138 & 1 34 03.3 & 30 46 36.6 & 3.0 & 70 & 0.11 & 0.03 & 0.23 & 0.04 & 0.20 & 0.02 & 0.047 & 0.004 & 0.011 & 0.001 \\
139 & 1 34 22.7 & 30 47 03.3 & 4.0 & 90 & 0.19 & 0.01 & 0.19 & 0.01 & 0.13 & 0.00 & 0.014 & 0.003 & 0.004 & 0.001 \\
140 & 1 34 41.5 & 30 47 14.7 & 5.3 & 92 & 0.22 & 0.02 & 0.42 & 0.03 & 0.27 & 0.01 & 0.075 & 0.005 & 0.013 & 0.001 \\
141 & 1 33 43.0 & 30 47 39.5 & 3.3 & 149 & 0.85 & 0.03 & 1.23 & 0.03 & 0.86 & 0.02 & 0.227 & 0.009 & 0.037 & 0.002 \\
142 & 1 33 32.8 & 30 47 22.0 & 3.5 & 126 & 0.23 & 0.02 & 0.41 & 0.02 & 0.35 & 0.01 & 0.040 & 0.006 & 0.010 & 0.001 \\
143 & 1 34 06.4 & 30 47 29.3 & 3.4 & 137 & 2.87 & 0.08 & 2.63 & 0.09 & 1.48 & 0.04 & 0.243 & 0.009 & 0.043 & 0.002 \\
144 & 1 33 51.2 & 30 47 39.0 & 3.2 & 148 & 0.32 & 0.01 & 0.52 & 0.02 & 0.41 & 0.01 & 0.033 & 0.006 & 0.017 & 0.002 \\
145 & 1 33 21.9 & 30 47 49.4 & 4.1 & 156 & 0.35 & 0.02 & 0.84 & 0.02 & 0.61 & 0.01 & 0.123 & 0.005 & 0.029 & 0.002 \\
146 & 1 33 28.0 & 30 47 42.0 & 3.8 & 78 & 0.23 & 0.01 & 0.34 & 0.03 & 0.24 & 0.01 & 0.044 & 0.005 & 0.007 & 0.001 \\
147 & 1 34 03.9 & 30 48 08.2 & 3.5 & 55 & 0.07 & 0.02 & 0.09 & 0.02 & 0.09 & 0.01 & 0.019 & 0.003 & 0 & 0.001 \\
148 & 1 33 59.2 & 30 48 56.2 & 3.7 & 175 & 2.84 & 0.05 & 3.25 & 0.04 & 2.21 & 0.02 & 0.489 & 0.013 & 0.111 & 0.004 \\
149 & 1 33 12.5 & 30 48 56.6 & 5.0 & 188 & 0.88 & 0.04 & 0.91 & 0.03 & 0.71 & 0.01 & 0.155 & 0.008 & 0.035 & 0.002 \\
150 & 1 33 36.3 & 30 49 04.6 & 4.0 & 197 & 0.71 & 0.03 & 1.25 & 0.03 & 0.97 & 0.01 & 0.219 & 0.007 & 0.050 & 0.002 \\
151 & 1 34 11.3 & 30 48 29.3 & 3.9 & 66 & 0.11 & 0.01 & 0.17 & 0.03 & 0.14 & 0.01 & 0.023 & 0.005 & 0.004 & 0.001 \\
152 & 1 34 19.4 & 30 48 28.0 & 4.2 & 65 & 0.20 & 0.03 & 0.29 & 0.04 & 0.19 & 0.02 & 0.038 & 0.006 & 0.006 & 0.001 \\
153 & 1 34 14.7 & 30 48 31.5 & 4.1 & 66 & 0.20 & 0.02 & 0.26 & 0.03 & 0.17 & 0.01 & 0.032 & 0.005 & 0.005 & 0.001 \\
154 & 1 34 22.1 & 30 48 38.9 & 4.4 & 54 & 0.18 & 0.04 & 0.25 & 0.04 & 0.15 & 0.02 & 0.032 & 0.007 & 0.003 & 0.001 \\
155 & 1 34 09.8 & 30 49 03.5 & 4.1 & 72 & 0.10 & 0.02 & 0.23 & 0.03 & 0.25 & 0.02 & 0.045 & 0.004 & 0.012 & 0.001 \\
156 & 1 33 13.4 & 30 50 28.3 & 5.4 & 83 & 0.17 & 0.01 & 0.18 & 0.01 & 0.14 & 0.01 & 0.034 & 0.003 & 0.005 & 0.001 \\
157 & 1 34 02.3 & 30 50 26.7 & 4.4 & 65 & 0.11 & 0.01 & 0.14 & 0.02 & 0.10 & 0.01 & 0.014 & 0.002 & 0.004 & 0.001 \\
158 & 1 33 36.3 & 30 50 50.2 & 4.6 & 95 & 0.48 & 0.02 & 0.46 & 0.02 & 0.27 & 0.01 & 0.034 & 0.004 & 0.012 & 0.001 \\
159 & 1 34 02.2 & 30 51 18.0 & 4.7 & 207 & 0.60 & 0.03 & 0.82 & 0.03 & 0.63 & 0.01 & 0.094 & 0.007 & 0.019 & 0.002 \\
160 & 1 34 16.5 & 30 52 06.4 & 5.4 & 280 & 4.25 & 0.06 & 4.91 & 0.06 & 3.11 & 0.02 & 0.600 & 0.013 & 0.149 & 0.004 \\
161 & 1 33 16.1 & 30 52 55.3 & 6.1 & 189 & 8.67 & 0.07 & 5.93 & 0.04 & 2.92 & 0.01 & 0.529 & 0.010 & 0.133 & 0.003 \\
162 & 1 34 02.1 & 30 52 37.9 & 5.2 & 127 & 0.24 & 0.01 & 0.27 & 0.02 & 0.28 & 0.00 & 0.036 & 0.006 & 0.016 & 0.001 \\
163 & 1 34 16.1 & 30 53 43.3 & 6.0 & 133 & 0.22 & 0.01 & 0.27 & 0.01 & 0.27 & 0.01 & 0.055 & 0.005 & 0.008 & 0.001 \\
164 & 1 34 04.1 & 30 54 34.9 & 6.0 & 235 & 0.97 & 0.03 & 1.05 & 0.03 & 1.12 & 0.01 & 0.229 & 0.009 & 0.044 & 0.003 \\
\hline
\end{tabular}
\end{table}

\end{landscape}

\section{SED Parameters}\label{app:sed_params}
\begin{landscape}
\begin{table}
\caption{Calculated dust and gas properties for the GMCs. Generally, errors given are 1$\sigma$ errors. However, in the case of an 3$\sigma$ upper limit, the reported value is 0 and the corresponding error is that 3$\sigma$ upper limit.}
\label{table:sed_parameters}
\begin{tabular}{cccccccccccc}
\hline \hline
GMC ID & T (K) & $\sigma_{\rm T}$ & log(${\rm M}_{\rm dust}$ [${\rm M}_\odot$]) & $\sigma_{\log({\rm M}_{\rm dust})}$ & log($L_{\rm TIR}$ [$L_\odot$]) & $\sigma_{\log(L_{\rm TIR})}$ & $L_{\rm CO}$ (K km s$^{-1}$) & $\sigma_{L_{\rm CO}}$ & $L_{\rm CO}/\sigma_{L_{\rm CO}} \times 10^x$ & $\Sigma_{\rm H\textsc{i}}$ (M$_\odot$ pc$^{-2}$) & $\sigma_{\Sigma_{\rm H\textsc{i}}}$ \\
\hline
0 & 20.95 & 0.54 & 3.76 & 0.05 & 5.54 & 0.02 & 1.95 & 0.15 & -1 & 5.17 & 0.18 \\
1 & 21.04 & 0.45 & 4.05 & 0.04 & 5.83 & 0.02 & 2.72 & 0.12 & -1 & 2.73 & 0.18 \\
2 & 18.13 & 0.42 & 3.43 & 0.04 & 4.86 & 0.03 & 9.47 & 1.60 & -2 & 1.87 & 0.21 \\
3 & 20.98 & 0.43 & 4.43 & 0.04 & 6.21 & 0.02 & 2.94 & 0.13 & -1 & 3.62 & 0.17 \\
4 & 19.58 & 0.41 & 3.92 & 0.04 & 5.54 & 0.02 & 1.25 & 0.10 & -1 & 4.52 & 0.34 \\
5 & 21.14 & 0.64 & 3.34 & 0.05 & 5.14 & 0.03 & 1.93 & 0.16 & -1 & 2.16 & 0.35 \\
6 & 20.44 & 0.50 & 3.74 & 0.04 & 5.46 & 0.02 & 2.74 & 0.85 & -2 & 1.83 & 0.10 \\
7 & 24.54 & 0.57 & 4.26 & 0.04 & 6.41 & 0.03 & 5.18 & 0.17 & -1 & 2.97 & 0.18 \\
8 & 22.76 & 0.61 & 3.38 & 0.04 & 5.35 & 0.03 & 4.03 & 1.26 & -2 & 2.95 & 0.16 \\
9 & 21.97 & 0.50 & 3.71 & 0.04 & 5.61 & 0.02 & 0 & 4.11 & -2 & 4.22 & 0.21 \\
10 & 25.55 & 0.78 & 2.98 & 0.05 & 5.23 & 0.03 & 2.18 & 0.28 & -1 & 0 & 0.60 \\
11 & 22.20 & 0.49 & 3.69 & 0.04 & 5.60 & 0.02 & 2.72 & 0.22 & -1 & 1.20 & 0.17 \\
12 & 20.20 & 0.61 & 3.51 & 0.05 & 5.20 & 0.03 & 9.63 & 1.75 & -2 & 1.09 & 0.21 \\
13 & 20.24 & 0.46 & 4.03 & 0.04 & 5.73 & 0.02 & 2.25 & 0.10 & -1 & 3.18 & 0.23 \\
14 & 24.50 & 0.61 & 4.04 & 0.04 & 6.20 & 0.03 & 3.34 & 0.33 & -1 & 1.87 & 0.26 \\
15 & 22.22 & 0.74 & 3.35 & 0.06 & 5.27 & 0.03 & 1.87 & 0.17 & -1 & 1.82 & 0.35 \\
16 & 20.29 & 0.43 & 4.02 & 0.04 & 5.72 & 0.02 & 1.93 & 0.12 & -1 & 3.68 & 0.21 \\
17 & 20.26 & 0.42 & 3.62 & 0.04 & 5.31 & 0.02 & 1.67 & 0.21 & -1 & 1.38 & 0.11 \\
18 & 24.56 & 0.60 & 3.89 & 0.04 & 6.05 & 0.03 & 2.01 & 0.21 & -1 & 3.01 & 0.42 \\
19 & 19.93 & 0.89 & 3.34 & 0.08 & 5.00 & 0.04 & 2.13 & 0.46 & -1 & 0 & 1.42 \\
20 & 22.57 & 0.93 & 2.95 & 0.07 & 4.90 & 0.05 & 3.17 & 0.28 & -1 & 1.51 & 0.28 \\
21 & 21.81 & 0.68 & 3.08 & 0.06 & 4.95 & 0.03 & 9.01 & 2.85 & -2 & 0 & 1.17 \\
22 & 21.47 & 0.58 & 3.54 & 0.05 & 5.38 & 0.03 & 2.48 & 0.19 & -1 & 2.04 & 0.19 \\
23 & 21.93 & 0.66 & 3.65 & 0.05 & 5.54 & 0.03 & 1.97 & 0.28 & -1 & 1.82 & 0.15 \\
24 & 28.33 & 0.85 & 3.84 & 0.05 & 6.34 & 0.03 & 1.57 & 0.14 & 0 & 0 & 3.19 \\
25 & 20.59 & 0.74 & 3.19 & 0.06 & 4.92 & 0.03 & 2.94 & 0.26 & -1 & 0 & 1.09 \\
26 & 19.01 & 0.86 & 3.57 & 0.08 & 5.11 & 0.04 & 6.97 & 0.93 & -1 & 0 & 1.53 \\
27 & 22.70 & 0.79 & 3.07 & 0.06 & 5.04 & 0.04 & 2.63 & 0.37 & -1 & 1.58 & 0.48 \\
28 & 24.69 & 0.94 & 3.10 & 0.06 & 5.27 & 0.04 & 4.20 & 0.55 & -1 & 2.38 & 0.44 \\
29 & 24.96 & 0.63 & 4.01 & 0.04 & 6.20 & 0.03 & 8.75 & 0.89 & -1 & 2.81 & 0.42 \\
30 & 23.39 & 1.23 & 3.15 & 0.08 & 5.19 & 0.06 & 7.19 & 1.13 & -1 & 0 & 2.89 \\
31 & 19.30 & 0.63 & 3.75 & 0.06 & 5.33 & 0.03 & 9.17 & 1.47 & -2 & 2.61 & 0.23 \\
32 & 20.90 & 0.54 & 3.50 & 0.05 & 5.27 & 0.03 & 1.80 & 0.19 & -1 & 1.72 & 0.53 \\
33 & 20.79 & 0.84 & 3.43 & 0.06 & 5.18 & 0.04 & 4.79 & 0.35 & -1 & 2.58 & 0.37 \\
34 & 24.99 & 0.71 & 3.92 & 0.05 & 6.12 & 0.03 & 8.52 & 0.32 & -1 & 2.97 & 0.26 \\
35 & 20.64 & 0.51 & 3.55 & 0.05 & 5.29 & 0.02 & 5.40 & 0.33 & -1 & 2.62 & 0.31 \\
36 & 28.84 & 1.51 & 2.58 & 0.09 & 5.12 & 0.05 & 2.08 & 0.48 & -1 & 0 & 1.67 \\
37 & 21.52 & 0.56 & 3.58 & 0.05 & 5.42 & 0.03 & 1.23 & 0.33 & -1 & 2.95 & 0.34 \\
38 & 27.23 & 1.19 & 3.41 & 0.06 & 5.81 & 0.04 & 3.69 & 0.59 & -1 & 0 & 2.62 \\
39 & 23.45 & 0.69 & 3.88 & 0.05 & 5.92 & 0.03 & 6.14 & 0.37 & -1 & 1.23 & 0.34 \\
40 & 24.33 & 0.80 & 3.69 & 0.06 & 5.82 & 0.03 & 4.91 & 0.32 & -1 & 2.24 & 0.40 \\
41 & 25.41 & 0.66 & 4.06 & 0.04 & 6.30 & 0.03 & 1.51 & 0.07 & 0 & 0 & 1.53 \\
42 & 21.65 & 0.83 & 3.50 & 0.06 & 5.35 & 0.04 & 2.96 & 0.29 & -1 & 0.99 & 0.20 \\
43 & 24.06 & 1.34 & 3.05 & 0.08 & 5.16 & 0.06 & 0 & 4.82 & -1 & 2.12 & 0.69 \\
\end{tabular}
\end{table}
\end{landscape}
\begin{landscape}
\begin{table}
\contcaption{Calculated dust and gas properties for the GMCs. Generally, errors given are 1$\sigma$ errors. However, in the case of an 3$\sigma$ upper limit, the reported value is 0 and the corresponding error is that 3$\sigma$ upper limit.}
\label{table:sed_parameters_b}
\begin{tabular}{cccccccccccc}
\hline \hline
GMC ID & T (K) & $\sigma_{\rm T}$ & log(${\rm M}_{\rm dust}$ [${\rm M}_\odot$]) & $\sigma_{\log({\rm M}_{\rm dust})}$ & log($L_{\rm TIR}$ [$L_\odot$]) & $\sigma_{\log(L_{\rm TIR})}$ & $L_{\rm CO}$ (K km s$^{-1}$) & $\sigma_{L_{\rm CO}}$ & $L_{\rm CO}/\sigma_{L_{\rm CO}} \times 10^x$ & $\Sigma_{\rm H\textsc{i}}$ (M$_\odot$ pc$^{-2}$) & $\sigma_{\Sigma_{\rm H\textsc{i}}}$ \\
\hline
44 & 23.31 & 0.49 & 4.10 & 0.04 & 6.13 & 0.02 & 2.11 & 0.13 & -1 & 2.97 & 0.17 \\
45 & 23.92 & 0.97 & 2.84 & 0.07 & 4.93 & 0.04 & 2.29 & 0.25 & -1 & 2.32 & 0.23 \\
46 & 21.08 & 0.60 & 3.41 & 0.06 & 5.20 & 0.03 & 1.64 & 0.19 & -1 & 0 & 1.51 \\
47 & 23.76 & 0.53 & 4.07 & 0.04 & 6.15 & 0.02 & 8.03 & 0.43 & -1 & 2.98 & 0.37 \\
48 & 24.21 & 0.70 & 3.98 & 0.05 & 6.10 & 0.03 & 4.30 & 0.34 & -1 & 3.88 & 0.44 \\
49 & 22.43 & 0.55 & 3.99 & 0.04 & 5.93 & 0.02 & 2.35 & 0.12 & -1 & 2.17 & 0.18 \\
50 & 22.44 & 0.59 & 3.67 & 0.05 & 5.62 & 0.03 & 0 & 3.66 & -2 & 4.21 & 0.33 \\
51 & 28.97 & 0.99 & 2.80 & 0.05 & 5.35 & 0.04 & 2.32 & 0.29 & -1 & 0 & 1.47 \\
52 & 19.79 & 0.51 & 3.70 & 0.05 & 5.34 & 0.02 & 3.83 & 0.25 & -1 & 0 & 1.06 \\
53 & 22.71 & 0.49 & 4.12 & 0.04 & 6.09 & 0.02 & 1.72 & 0.12 & -1 & 5.06 & 0.20 \\
54 & 18.46 & 0.36 & 4.43 & 0.04 & 5.90 & 0.02 & 3.31 & 0.11 & -1 & 4.23 & 0.14 \\
55 & 24.88 & 0.87 & 3.27 & 0.06 & 5.46 & 0.04 & 0 & 18.63 & -2 & 2.64 & 0.22 \\
56 & 27.34 & 0.75 & 4.07 & 0.04 & 6.48 & 0.03 & 7.45 & 0.46 & -1 & 1.92 & 0.23 \\
57 & 21.00 & 0.48 & 3.74 & 0.04 & 5.52 & 0.02 & 7.75 & 1.24 & -2 & 2.69 & 0.29 \\
58 & 25.65 & 0.62 & 3.88 & 0.04 & 6.14 & 0.03 & 8.60 & 0.41 & -1 & 1.93 & 0.46 \\
59 & 23.96 & 1.18 & 3.36 & 0.07 & 5.46 & 0.06 & 9.48 & 0.85 & -1 & 1.72 & 0.39 \\
60 & 20.07 & 0.49 & 4.05 & 0.05 & 5.72 & 0.02 & 1.07 & 0.10 & -1 & 3.48 & 0.16 \\
61 & 25.94 & 0.91 & 3.66 & 0.06 & 5.95 & 0.04 & 9.09 & 0.56 & -1 & 2.31 & 0.44 \\
62 & 29.58 & 2.05 & 2.91 & 0.09 & 5.51 & 0.08 & 6.08 & 0.87 & -1 & 0 & 2.77 \\
63 & 26.67 & 0.80 & 3.72 & 0.05 & 6.08 & 0.03 & 1.19 & 0.28 & -1 & 2.73 & 0.29 \\
64 & 24.53 & 0.98 & 2.95 & 0.07 & 5.10 & 0.04 & 0 & 15.27 & -2 & 3.42 & 0.82 \\
65 & 27.88 & 1.15 & 3.10 & 0.06 & 5.56 & 0.04 & 5.26 & 0.93 & -1 & 0 & 1.92 \\
66 & 19.68 & 0.71 & 3.82 & 0.06 & 5.45 & 0.04 & 9.62 & 0.66 & -1 & 4.33 & 0.36 \\
67 & 27.03 & 0.90 & 3.27 & 0.05 & 5.65 & 0.04 & 5.18 & 0.60 & -1 & 0 & 1.18 \\
68 & 23.87 & 0.90 & 2.99 & 0.07 & 5.07 & 0.04 & 3.25 & 0.50 & -1 & 1.27 & 0.30 \\
69 & 22.23 & 0.48 & 3.50 & 0.04 & 5.42 & 0.02 & 1.46 & 0.19 & -1 & 3.24 & 0.27 \\
70 & 18.97 & 0.44 & 3.90 & 0.04 & 5.44 & 0.02 & 3.28 & 0.21 & -1 & 1.78 & 0.20 \\
71 & 21.49 & 0.77 & 3.57 & 0.06 & 5.41 & 0.04 & 3.70 & 0.33 & -1 & 1.99 & 0.43 \\
72 & 25.14 & 1.49 & 2.60 & 0.10 & 4.81 & 0.06 & 5.39 & 1.27 & -1 & 0 & 1.02 \\
73 & 26.85 & 1.44 & 2.67 & 0.08 & 5.04 & 0.06 & 1.12 & 0.17 & 0 & 0 & 1.57 \\
74 & 17.57 & 0.54 & 3.86 & 0.06 & 5.22 & 0.03 & 3.61 & 0.17 & -1 & 2.05 & 0.26 \\
75 & 25.86 & 1.04 & 3.33 & 0.06 & 5.61 & 0.04 & 4.29 & 0.50 & -1 & 0 & 1.49 \\
76 & 25.54 & 0.75 & 3.17 & 0.05 & 5.42 & 0.03 & 6.74 & 0.40 & -1 & 1.05 & 0.28 \\
77 & 24.89 & 0.55 & 4.55 & 0.04 & 6.74 & 0.02 & 1.12 & 0.03 & 0 & 4.87 & 0.20 \\
78 & 27.77 & 0.73 & 4.20 & 0.04 & 6.65 & 0.03 & 1.02 & 0.04 & 0 & 3.73 & 0.23 \\
79 & 21.27 & 0.45 & 4.27 & 0.04 & 6.08 & 0.02 & 3.52 & 0.18 & -1 & 4.33 & 0.15 \\
80 & 26.08 & 0.65 & 3.87 & 0.04 & 6.17 & 0.03 & 1.66 & 0.20 & -1 & 1.47 & 0.28 \\
81 & 27.39 & 1.29 & 3.25 & 0.07 & 5.67 & 0.05 & 4.98 & 0.90 & -1 & 3.33 & 0.68 \\
82 & 23.04 & 0.54 & 4.03 & 0.04 & 6.03 & 0.02 & 2.81 & 0.14 & -1 & 2.14 & 0.18 \\
83 & 28.94 & 0.96 & 3.27 & 0.05 & 5.82 & 0.04 & 0 & 13.95 & -2 & 1.47 & 0.19 \\
84 & 26.67 & 0.85 & 3.62 & 0.05 & 5.97 & 0.04 & 4.77 & 0.44 & -1 & 2.66 & 0.34 \\
85 & 24.28 & 0.58 & 4.08 & 0.04 & 6.21 & 0.03 & 6.98 & 0.34 & -1 & 1.83 & 0.38 \\
86 & 24.54 & 0.63 & 4.05 & 0.04 & 6.21 & 0.03 & 1.25 & 0.05 & 0 & 3.37 & 0.40 \\
87 & 24.53 & 1.96 & 2.95 & 0.11 & 5.10 & 0.09 & 1.80 & 0.35 & 0 & 1.03 & 0.25 \\
\end{tabular}
\end{table}
\end{landscape}
\begin{landscape}
\begin{table}
\contcaption{Calculated dust and gas properties for the GMCs. Generally, errors given are 1$\sigma$ errors. However, in the case of an 3$\sigma$ upper limit, the reported value is 0 and the corresponding error is that 3$\sigma$ upper limit.}
\label{table:sed_parameters_c}
\begin{tabular}{cccccccccccc}
\hline \hline
GMC ID & T (K) & $\sigma_{\rm T}$ & log(${\rm M}_{\rm dust}$ [${\rm M}_\odot$]) & $\sigma_{\log({\rm M}_{\rm dust})}$ & log($L_{\rm TIR}$ [$L_\odot$]) & $\sigma_{\log(L_{\rm TIR})}$ & $L_{\rm CO}$ (K km s$^{-1}$) & $\sigma_{L_{\rm CO}}$ & $L_{\rm CO}/\sigma_{L_{\rm CO}} \times 10^x$ & $\Sigma_{\rm H\textsc{i}}$ (M$_\odot$ pc$^{-2}$) & $\sigma_{\Sigma_{\rm H\textsc{i}}}$ \\
\hline
88 & 27.26 & 0.86 & 3.72 & 0.05 & 6.12 & 0.04 & 8.33 & 0.60 & -1 & 2.04 & 0.41 \\
89 & 19.09 & 0.46 & 3.56 & 0.05 & 5.12 & 0.02 & 6.81 & 1.41 & -2 & 2.31 & 0.33 \\
90 & 25.25 & 1.02 & 3.06 & 0.06 & 5.29 & 0.04 & 4.77 & 0.94 & -1 & 0 & 1.93 \\
91 & 18.63 & 0.51 & 3.74 & 0.05 & 5.24 & 0.03 & 1.93 & 0.17 & -1 & 1.40 & 0.24 \\
92 & 19.84 & 0.45 & 3.47 & 0.04 & 5.11 & 0.02 & 9.80 & 2.19 & -2 & 0.66 & 0.18 \\
93 & 17.02 & 0.52 & 3.97 & 0.06 & 5.25 & 0.03 & 7.69 & 0.67 & -1 & 2.71 & 0.32 \\
94 & 23.57 & 0.64 & 3.63 & 0.04 & 5.69 & 0.03 & 8.26 & 0.62 & -1 & 2.73 & 0.35 \\
95 & 19.92 & 1.19 & 3.23 & 0.09 & 4.88 & 0.06 & 4.12 & 0.44 & -1 & 0 & 1.12 \\
96 & 19.59 & 0.37 & 4.47 & 0.04 & 6.08 & 0.02 & 2.87 & 0.08 & -1 & 4.82 & 0.12 \\
97 & 30.75 & 0.95 & 3.74 & 0.04 & 6.43 & 0.04 & 4.21 & 0.31 & -1 & 0 & 1.34 \\
98 & 23.28 & 1.60 & 3.22 & 0.10 & 5.25 & 0.07 & 4.94 & 0.92 & -1 & 1.69 & 0.53 \\
99 & 23.59 & 0.57 & 4.26 & 0.04 & 6.32 & 0.03 & 8.14 & 0.43 & -1 & 4.42 & 0.17 \\
100 & 29.64 & 0.80 & 4.01 & 0.04 & 6.62 & 0.03 & 1.84 & 0.06 & 0 & 3.62 & 0.32 \\
101 & 21.33 & 1.01 & 3.17 & 0.08 & 4.99 & 0.05 & 4.85 & 0.94 & -1 & 1.75 & 0.46 \\
102 & 25.70 & 0.98 & 3.32 & 0.06 & 5.59 & 0.04 & 5.23 & 0.65 & -1 & 2.96 & 0.43 \\
103 & 20.51 & 0.85 & 3.37 & 0.07 & 5.10 & 0.04 & 1.98 & 0.18 & -1 & 1.28 & 0.10 \\
104 & 23.29 & 0.83 & 3.62 & 0.06 & 5.65 & 0.04 & 6.48 & 1.03 & -1 & 1.13 & 0.21 \\
105 & 29.54 & 0.80 & 4.41 & 0.04 & 7.01 & 0.03 & 6.11 & 0.17 & -1 & 2.92 & 0.35 \\
106 & 21.87 & 0.47 & 3.66 & 0.04 & 5.54 & 0.02 & 1.23 & 0.15 & -1 & 0.70 & 0.18 \\
107 & 25.20 & 0.64 & 3.76 & 0.04 & 5.97 & 0.03 & 1.79 & 0.49 & -1 & 4.95 & 0.47 \\
108 & 23.19 & 0.51 & 3.17 & 0.04 & 5.19 & 0.03 & 0 & 10.52 & -2 & 2.91 & 0.35 \\
109 & 25.66 & 1.19 & 3.21 & 0.07 & 5.47 & 0.05 & 3.48 & 0.66 & -1 & 0 & 1.36 \\
110 & 24.95 & 0.67 & 3.84 & 0.04 & 6.03 & 0.03 & 4.33 & 0.22 & -1 & 0 & 1.08 \\
111 & 19.55 & 0.59 & 3.64 & 0.05 & 5.25 & 0.03 & 3.78 & 0.30 & -1 & 1.49 & 0.26 \\
112 & 20.90 & 0.63 & 3.88 & 0.06 & 5.65 & 0.03 & 7.03 & 1.11 & -2 & 1.40 & 0.19 \\
113 & 26.49 & 0.73 & 3.57 & 0.04 & 5.91 & 0.03 & 7.32 & 0.53 & -1 & 0 & 1.45 \\
114 & 20.13 & 0.56 & 3.65 & 0.05 & 5.33 & 0.03 & 2.15 & 0.18 & -1 & 0 & 1.14 \\
115 & 26.87 & 1.02 & 2.99 & 0.06 & 5.36 & 0.05 & 0 & 7.63 & -2 & 0 & 1.10 \\
116 & 21.27 & 0.64 & 3.75 & 0.05 & 5.56 & 0.03 & 7.92 & 0.38 & -1 & 1.00 & 0.20 \\
117 & 21.50 & 0.67 & 3.36 & 0.05 & 5.20 & 0.03 & 2.39 & 0.25 & -1 & 1.69 & 0.22 \\
118 & 24.73 & 1.04 & 3.04 & 0.07 & 5.21 & 0.05 & 6.38 & 0.81 & -1 & 0 & 1.64 \\
119 & 22.36 & 1.33 & 3.20 & 0.09 & 5.13 & 0.06 & 3.59 & 0.27 & -1 & 0 & 0.68 \\
120 & 20.46 & 0.87 & 3.13 & 0.07 & 4.85 & 0.04 & 5.17 & 0.70 & -1 & 0 & 1.44 \\
121 & 18.59 & 0.47 & 3.49 & 0.05 & 4.98 & 0.03 & 1.76 & 0.34 & -1 & 2.49 & 0.43 \\
122 & 23.88 & 0.58 & 3.93 & 0.04 & 6.02 & 0.03 & 4.79 & 0.18 & -1 & 2.43 & 0.19 \\
123 & 23.45 & 0.97 & 3.35 & 0.07 & 5.39 & 0.04 & 2.75 & 0.18 & -1 & 0 & 2.59 \\
124 & 28.05 & 1.38 & 2.92 & 0.07 & 5.39 & 0.05 & 3.37 & 0.38 & -1 & 3.04 & 0.69 \\
125 & 20.38 & 0.92 & 3.51 & 0.07 & 5.22 & 0.05 & 4.89 & 0.47 & -1 & 0 & 1.80 \\
126 & 28.10 & 0.73 & 3.95 & 0.04 & 6.43 & 0.03 & 3.76 & 0.17 & -1 & 0 & 0.87 \\
127 & 18.27 & 0.53 & 3.48 & 0.06 & 4.93 & 0.03 & 2.76 & 0.67 & -1 & 1.27 & 0.18 \\
128 & 19.34 & 0.50 & 3.54 & 0.05 & 5.12 & 0.03 & 1.26 & 0.19 & -1 & 1.28 & 0.18 \\
129 & 20.26 & 0.63 & 3.49 & 0.06 & 5.18 & 0.03 & 5.20 & 0.52 & -1 & 4.01 & 0.67 \\
130 & 25.76 & 0.59 & 4.18 & 0.04 & 6.46 & 0.03 & 2.91 & 0.15 & -1 & 2.30 & 0.22 \\
131 & 25.02 & 1.06 & 3.12 & 0.07 & 5.32 & 0.04 & 2.54 & 0.57 & -1 & 3.02 & 0.65 \\
\end{tabular}
\end{table}
\end{landscape}
\begin{landscape}
\begin{table}
\contcaption{Calculated dust and gas properties for the GMCs. Generally, errors given are 1$\sigma$ errors. However, in the case of an 3$\sigma$ upper limit, the reported value is 0 and the corresponding error is that 3$\sigma$ upper limit.}
\label{table:sed_parameters_d}
\begin{tabular}{cccccccccccc}
\hline \hline
GMC ID & T (K) & $\sigma_{\rm T}$ & log(${\rm M}_{\rm dust}$ [${\rm M}_\odot$]) & $\sigma_{\log({\rm M}_{\rm dust})}$ & log($L_{\rm TIR}$ [$L_\odot$]) & $\sigma_{\log(L_{\rm TIR})}$ & $L_{\rm CO}$ (K km s$^{-1}$) & $\sigma_{L_{\rm CO}}$ & $L_{\rm CO}/\sigma_{L_{\rm CO}} \times 10^x$ & $\Sigma_{\rm H\textsc{i}}$ (M$_\odot$ pc$^{-2}$) & $\sigma_{\Sigma_{\rm H\textsc{i}}}$ \\
\hline
132 & 28.43 & 0.99 & 3.51 & 0.05 & 6.02 & 0.04 & 3.35 & 0.52 & -1 & 0 & 2.03 \\
133 & 21.75 & 0.51 & 4.32 & 0.04 & 6.18 & 0.02 & 3.44 & 0.15 & -1 & 3.63 & 0.15 \\
134 & 22.21 & 0.47 & 4.33 & 0.04 & 6.25 & 0.02 & 8.79 & 0.32 & -1 & 2.88 & 0.35 \\
135 & 29.56 & 0.79 & 4.98 & 0.04 & 7.58 & 0.03 & 1.50 & 0.01 & 0 & 4.00 & 0.26 \\
136 & 20.23 & 0.88 & 3.25 & 0.08 & 4.94 & 0.04 & 4.79 & 0.64 & -1 & 0 & 1.19 \\
137 & 20.66 & 0.43 & 3.74 & 0.04 & 5.49 & 0.02 & 1.46 & 0.17 & -1 & 1.76 & 0.20 \\
138 & 18.79 & 1.03 & 3.55 & 0.09 & 5.07 & 0.06 & 6.68 & 0.69 & -1 & 1.36 & 0.33 \\
139 & 24.38 & 0.69 & 3.01 & 0.05 & 5.15 & 0.03 & 0 & 3.74 & -2 & 1.57 & 0.22 \\
140 & 19.96 & 0.66 & 3.64 & 0.06 & 5.30 & 0.03 & 3.72 & 0.23 & -1 & 2.74 & 0.75 \\
141 & 21.00 & 0.45 & 4.06 & 0.04 & 5.84 & 0.02 & 4.26 & 0.33 & -1 & 2.80 & 0.21 \\
142 & 20.29 & 0.49 & 3.63 & 0.05 & 5.33 & 0.03 & 3.03 & 0.22 & -1 & 2.00 & 0.23 \\
143 & 26.22 & 0.70 & 4.02 & 0.04 & 6.33 & 0.03 & 6.77 & 0.60 & -1 & 2.71 & 0.44 \\
144 & 21.40 & 0.43 & 3.62 & 0.04 & 5.44 & 0.02 & 1.65 & 0.15 & -1 & 3.02 & 0.13 \\
145 & 18.37 & 0.36 & 4.09 & 0.04 & 5.56 & 0.02 & 0 & 4.76 & -2 & 1.97 & 0.08 \\
146 & 21.54 & 0.53 & 3.45 & 0.04 & 5.29 & 0.03 & 1.36 & 0.38 & -1 & 2.60 & 0.46 \\
147 & 23.01 & 1.35 & 2.77 & 0.09 & 4.78 & 0.07 & 0 & 23.41 & -2 & 0 & 1.16 \\
148 & 22.89 & 0.51 & 4.35 & 0.04 & 6.34 & 0.02 & 1.18 & 0.02 & 0 & 2.31 & 0.13 \\
149 & 22.84 & 0.63 & 3.84 & 0.05 & 5.82 & 0.03 & 2.01 & 0.19 & -1 & 3.33 & 0.28 \\
150 & 19.43 & 0.40 & 4.21 & 0.04 & 5.81 & 0.02 & 2.24 & 0.15 & -1 & 1.33 & 0.18 \\
151 & 20.80 & 0.80 & 3.21 & 0.07 & 4.97 & 0.04 & 7.30 & 0.83 & -1 & 0 & 2.33 \\
152 & 22.37 & 1.03 & 3.29 & 0.07 & 5.23 & 0.05 & 0 & 27.59 & -2 & 0 & 1.36 \\
153 & 22.95 & 0.90 & 3.21 & 0.06 & 5.20 & 0.04 & 3.62 & 0.75 & -1 & 0 & 2.15 \\
154 & 24.10 & 1.46 & 3.07 & 0.08 & 5.18 & 0.07 & 6.13 & 1.28 & -1 & 0 & 2.26 \\
155 & 17.92 & 0.66 & 3.67 & 0.07 & 5.07 & 0.04 & 2.48 & 0.22 & 0 & 1.75 & 0.50 \\
156 & 22.72 & 0.78 & 3.13 & 0.06 & 5.10 & 0.03 & 1.73 & 0.26 & -1 & 0 & 1.22 \\
157 & 23.15 & 0.80 & 2.93 & 0.06 & 4.95 & 0.03 & 0 & 19.48 & -2 & 0 & 1.50 \\
158 & 25.61 & 0.77 & 3.30 & 0.05 & 5.55 & 0.03 & 6.66 & 1.73 & -2 & 0 & 1.15 \\
159 & 21.66 & 0.50 & 3.84 & 0.04 & 5.70 & 0.02 & 1.26 & 0.16 & -1 & 2.34 & 0.14 \\
160 & 23.20 & 0.49 & 4.49 & 0.04 & 6.51 & 0.02 & 2.90 & 0.06 & -1 & 3.33 & 0.11 \\
161 & 30.45 & 0.86 & 4.17 & 0.04 & 6.84 & 0.03 & 2.05 & 0.11 & -1 & 2.34 & 0.28 \\
162 & 21.36 & 0.51 & 3.48 & 0.05 & 5.30 & 0.02 & 9.02 & 1.45 & -2 & 1.72 & 0.14 \\
163 & 20.60 & 0.57 & 3.52 & 0.05 & 5.26 & 0.03 & 8.50 & 0.99 & -2 & 2.17 & 0.15 \\
164 & 21.56 & 0.52 & 4.05 & 0.04 & 5.89 & 0.02 & 1.34 & 0.07 & -1 & 3.14 & 0.10 \\
\hline
\end{tabular}
\end{table}

\end{landscape}


\bsp	
\label{lastpage}
\end{document}